\documentclass[reprint,twocolumn,amsmath,amssymb,aps,pra,longbibliography,floatfix]{revtex4-1}
\usepackage{physics}
\usepackage{amsmath}
\usepackage{cases}
\usepackage{url}
\usepackage{natbib}
\usepackage{textcase}
\usepackage{amssymb}
\usepackage{graphicx}
\usepackage{bm} 
\usepackage{times}
\usepackage{float}
\usepackage{multirow,microtype,color,relsize}
\usepackage{subfigure}
\usepackage[breaklinks=true]{hyperref}
\usepackage[utf8]{inputenc}
\usepackage[english]{babel}
\hypersetup{colorlinks=true,linkcolor=blue,citecolor=blue}
\hypersetup{linktocpage}
\usepackage{CJKutf8}
\usepackage{breakcites}
\usepackage{float}
\usepackage[dvipsnames]{xcolor}
\definecolor{mypine}{RGB}{1, 121, 111}

\begin{document}
\title{Understanding disorder in Silicon quantum computing platforms: Scattering mechanisms in Si/SiGe quantum wells}
\author{Yi Huang}
\author{Sankar Das Sarma}
%\email[Corresponding author: ]{huan1756@umn.edu}
\affiliation{Condensed Matter Theory Center and Joint Quantum Institute, Department of Physics, University of Maryland, College Park, Maryland 20742, USA}

\begin{abstract}	
Motivated by recent experiments on Si/SiGe quantum wells with a co-design of high electron mobility and large valley splitting [B. Paquelet Wuetz, \textit{et al.}, Nature Communications \textbf{14}, 1385 (2023); D. D. Esposti, \textit{et al.}, arXiv:2309.02832], suitable for a Si-based spin qubit quantum computing platform, we examine the role of disorder by theoretically calculating mobility and quantum mobility from various scattering mechanisms and their dependence on the electron density.
At low electron densities $n_e < 4 \times 10^{11}$ cm$^{-2}$, we find that mobility is limited by remote Coulomb impurities in the capping layer, whereas interface roughness becomes the significant limiting factor at higher densities.
We also find that alloy disorder scattering is not a limiting mechanism in the reported high-mobility structures.
We estimate the critical density of the disorder-driven low-density metal-insulator transition using the Anderson-Ioffe-Regel localization criterion and qualitatively explain the breakdown of the Boltzmann-Born theory at low densities.
We also estimate the critical density by considering inhomogeneous density fluctuations induced by long-range Coulomb disorder in the system, and find a larger critical density compared to the one obtained from the Anderson-Ioffe-Regel criterion.
For quantum mobility, our calculation suggests that remote and distant background impurities are likely the limiting scattering sources across all density.
Future measurements of quantum mobility should provide more information on the distribution of background impurities inside the SiGe barriers.
Moreover, we extend our theoretical analysis to the effect of quantum degeneracy on transport properties and predict the mobility and the critical density for the metal-insulator transition in spin-polarized high-mobility structures under an external parallel magnetic field.

\end{abstract}
\maketitle
%\onecolumngrid
\section{Introduction}
For spin qubits in gate-defined silicon quantum dots, it is important to minimize disorder potential fluctuations to better control charging energies and tunneling between quantum dots, which requires high mobility of the hosted two-dimensional electron gas (2DEG)~\cite{Vandersypen2017,Scappucci2021,Burkard:2023}.
Moreover, background charge noise, necessarily associated with unintentional random impurities in the environment, is the primary mechanism causing charge noise induced decoherence in semiconductor qubits, and therefore, a comprehensive understanding of the background impurity disorder is essential for good qubit performance~\cite{Xuedong:2006,Culcer:2009,Kestner:2021,Burkard:2023}.
Additionally, a large valley splitting energy is important to limit leakage from the computational Hilbert space in the lowest conduction band, which is crucial for maintaining high-fidelity qubit operations~\cite{Culcer:2010,Buterakos:2021,Zwanenburg:2013,Hu:2014,Tagliaferri:2018,Zwerver:2023,Burkard:2023}.
Recent advancements in isotopically purified $^{28}$ Si/SiGe quantum wells (QWs) at a low temperature of 1.7 K have shown promising improvements in channel static disorder, as indicated by an enhanced mobility $3 \times 10^5$~cm$^2$/Vs and a smaller critical density of metal-insulator transition $n_c = 7\times 10^{10}$ cm$^{-2}$, while maintaining a relatively high mean valley splitting $0.2$ meV~\cite{Wuetz:2023,Esposti:2023}.
(Isotopic purification suppresses nuclear spin noise for semiconductor based spin qubits in the Si platform, providing long spin coherence times and enabling high-fidelity gate operations~\cite{Witzel:2010}.)
In comparison, previous experimental reports of large valley splitting $> 0.2$~meV in Si/SiGe QWs showed 5 times lower mobility $< 6 \times 10^4$~cm$^2$/Vs~\cite{Borselli:2011,Hollmann:2020,McJunkin:2022}.
On the other hand, high mobility of $6.5 \times 10^5$~cm$^2$/Vs was reported in conventional Si/SiGe heterostructures but with low valley splitting $35-70$~$\mu$eV\cite{Petta:2017}.
Therefore, the co-design of high electron mobility and large valley splitting calls for reinvestigating the scattering sources inside the Si/SiGe QWs reported in Refs.~\cite{Wuetz:2023,Esposti:2023}. 
This is the goal of our theoretical work, particularly in the context of the importance of the Si/Ge quantum computing platform.

In this study, we employ the Boltzmann-Born transport theory~\cite{Ando_review:1982} to theoretically investigate both transport and quantum mobility ($\mu$ and $\mu_{q}$).
Our focus is on quantifying the effects of various scattering mechanisms, namely remote impurities at the semiconductor-oxide interface (RI), background impurities (BI), interface roughness (IR), and alloy disorder (AD).
By analyzing experimental mobility data from 5 nm and 7 nm Si/SiGe quantum wells reported in Refs.~\cite{Wuetz:2023,Esposti:2023}, we find that RI predominantly limits $\mu$ at electron densities below $4 \times 10^{11}$ cm$^{-2}$, while IR becomes the limiting factor at higher densities ranging from $4-6 \times 10^{11}$ cm$^{-2}$.
Despite limited experimental data on quantum mobility ($\mu_q =3.0 \pm 0.5 \times 10^4$~cm$^2$/Vs at the highest density $6 \times 10^{11}$ cm$^{-2}$~\cite{private_comm}), our theoretical models indicate that $\mu_{q}$ is likely constrained by RI and distant BI scattering across all electron densities.
Since $\mu_q$ is much more sensitive to RI and distant BI scattering compared to $\mu$~\cite{Stern:1985}, we need more experimental data on $\mu_q$ across a wider range of densities to better understand the distribution for BI in SiGe barriers. 
We also show that AD is quantitatively irrelevant in all experimental density ranges for these high-mobility structures.
We address the limitations of the Boltzmann-Born theory at low electron densities $n_e \lesssim 2 \times 10^{11}$ cm$^{-2}$ with considerable disagreement between the theory and experimental data, although our theory explains the higher density data quantitatively.
A possible explanation for this discrepancy at low densities is that the 2DEG is broken into inhomogeneous electron puddles separated by long-range Coulomb disorder potential barriers, so that both the perturbative Born approximation and linear screening are no longer applicable.
The system eventually undergoes a metal-insulator transition (MIT) at a critical density $n_e = n_c$, where transport in the low-density $n_e \geq n_c$ region should be described by percolation through charged puddles~\cite{Thouless:1971,Kirkpatrick:1973,Shklovskii:1975,DasSarma:2005579,Tracy:2009,Manfra:2007,Qiuzi:2013,Tracy:2014}.
This MIT is, in fact, a crossover (happening sharply over a narrow density range around $n_c$ because of the rapid failure of screening of the long-range Coulomb disorder) and not a quantum phase transition~\cite{DasSarma:2005579,Tracy:2009}. 
We estimate $n_c$ in the experimental systems~\cite{Wuetz:2023,Esposti:2023} by using the Anderson-Ioffe-Regel (AIR) criterion~\cite{anderson1958,ioffe1960}, although the failure of the Boltzmann-Born theory at low densities indicates that the critical density estimated from the AIR criterion should be smaller than the actual critical density (percolation threshold) observed in experiments and should be used as a lower bound estimate of the percolation threshold. 
We also estimate $n_c$ by considering inhomogeneous density fluctuations induced by long-range Coulomb disorder in the system~\cite{Efros:19881019,Pikus:1989}, and find a larger $n_c$ compared to the one obtained from the AIR criterion.
Moreover, we extend our theoretical analysis to the effect of quantum degeneracy on transport properties and predict the mobility and the critical density for the metal-insulator transition in spin-polarized high-mobility structures, which can be tested in future experiments by applying an external parallel magnetic field.

\begin{figure}[t]
    \centering
    \includegraphics[width = 0.7\linewidth]{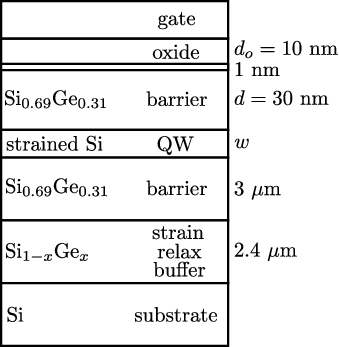}%mu_low_ne0.eps
    \caption{A schematic illustration of the high-mobility Si/SiGe heterostructure reported in Refs.~\cite{Wuetz:2023,Esposti:2023}, where $w$ is the thickness of the strained Si quantum well. The oxide and the top Si$_{0.69}$Ge$_{0.31}$ barrier is separated by a Si capping layer of thickness 1 nm.}
    \label{fig:schematics}
\end{figure}

The Si/SiGe heterostructures reported in Refs.~\cite{Wuetz:2023,Esposti:2023} consist of a strained Si QW of width $w$ surrounded by a Si$_{0.69}$Ge$_{0.31}$ upper barrier of thickness $d = 30$ nm and a thick strained relaxed Si$_{0.69}$Ge$_{0.31}$ bottom barrier of several $\mu$m. 
The upper barrier is passivated with a 1 nm Si capping layer. 
A metallic top gate is separated from the capping layer by a $d_o=10$ nm oxide dielectric spacer.
The electron density $n_e$ is tuned by the gate up to $ 6\times 10^{11}$ cm$^{-2}$.
The bottom barrier is placed on top of a step-graded Si$_{1-x}$Ge$_{x}$ strain-relax buffer layer, which has a decreasing Ge concentration $x$ until it reaches the Si substrate with $x=0$.
A schematic illustration of the Si/SiGe heterostructure is shown in Fig.~\ref{fig:schematics}.
In our calculations, we model the possible disorder sources as follows.
There are uniform background charged impurities outside (inside) the QW with 3D concentration $N_1$ ($N_2$).
The remote charged impurities at the semiconductor-oxide interface have a 2D concentration $n_{r}$.
Because the top gate is relatively close to the semiconductor-oxide interface, we take into account the gate screening when we calculate the remote and background charged impurity scattering.
The interface roughness between the Si QW and the Si$_{0.69}$Ge$_{0.31}$ barrier is characterized by a typical height $\Delta$ and a lateral size $\Lambda$.
The conduction band offset from the twofold valley degenerate conduction band $\Delta_2$ in the strained Si to the sixfold valley degenerate conduction band $\Delta_6$ in the relaxed Si$_{0.69}$Ge$_{0.31}$ is taken to be $V_0 =180$ meV~\cite{Stern:1992,Burkard:2023}.
The 2DEG wave function penetrates the barriers and scatters from the alloy disorder.
%Due to this finite potential confinement, the 2DEG wave function leaks into the Si$_{0.69}$Ge$_{0.31}$ barrier and scatters from the alloy disorder inside the barrier.
Scattering mechanisms considered in our theory are the only low-temperature scattering mechanisms operational in high quality SiGe 2D structures, where the crystal quality is good enough to rule out short-range scattering by atomic point defects and vacancies~\cite{crystal_quality}, but random charged impurities are invariably present in the system along with interface roughness and alloy disorder.  
Since our interest is in understanding the low-temperature mobility, we neglect all phonon scattering and restrict our calculations to $T=0$.
We believe that our comprehensive transport theory is quantitatively accurate at high carrier densities, and poorer at lower densities although it should remain qualitatively (and semiquantitatively) valid at lower densities above the critical density for the 2D MIT.

The rest of this paper is structured as follows.
Section~\ref{sec:mu_muq} discusses the general formalism of the Boltzmann transport theory to calculate the mobility and quantum mobility. We evaluate the BI and RI scattering rates using the delta-layer 2DEG approximation in Section~\ref{sec:delta}. The infinite potential well approximation together with the local field correction, applied to RI, BI, and IR scattering, is discussed in Section~\ref{sec:inf}, while Section~\ref{sec:finite} employs a finite potential well approximation to calculate the AD scattering. Gate-screened charged impurity scattering is explored in Section~\ref{sec:gate}. In Section~\ref{sec:mit}, we estimate the critical densities of MIT using the Anderson-Ioffe-Regel (AIR) criterion and qualitatively explain the breakdown of the Boltzmann-Born theory at low densities. We mention that the AIR criterion is often referred to as the Mott-Ioffe-Regel (MIR or IRM) or just the Ioffe-Regel criterion, but we believe that AIR is a more appropriate terminology here since our criterion is specific to disorder induced incoherent scattering associated with destructive interference as envisioned by Anderson rather than the Mott criterion involving the transport mean free path being equal to lattice spacing, which plays no role in doped semiconductors of interest to us. We also estimate the critical density by considering inhomogeneous density fluctuations induced by long-range Coulomb disorder in the system and compare the result to the AIR critical density.
Section~\ref{sec:valley} examines the effect of spin/valley degeneracy on the mobility and the AIR critical density.
Finally, we conclude and summarize in Section~\ref{sec:conclusion}.

\section{Mobility and quantum mobility}
\label{sec:mu_muq}
In this section, we apply the Boltzmann transport theory to calculate the mobility and quantum mobility at $T=0$~\cite{Ando_review:1982,Hwang:1999,Hwang_screening:2005,Hwang:2007,Hwang:2013a_valley,Hwang:2014,Binhui:2015}. 
The transport mobility is given by $\mu = e \tau / m^{\star}$, and the conductivity is $e n_e \mu$, where $n_e$ is the 2DEG concentration, $m^\star = 0.19\, m_0$ is the transverse effective mass of an electron in Si, $m_0$ is the free electron mass, 
\begin{align}\label{eq:1_tau}
	\frac{1}{\tau} =\frac{4m^{\star}}{\pi \hbar^3} \int \limits_0^{2k_F}\frac{dq}{\sqrt{4k_F^2 - q^2}} \qty(\frac{q}{2k_F})^2 \ev{\abs{U(q)}^2}\,,
\end{align}
is the transport scattering rate, and $U(q)$ is the screened potential of a given scattering source.
$k_F = \sqrt{4\pi n_e/g}$ is the Fermi wave vector.
$g=g_s g_v$ is the total degeneracy of electrons, where $g_s$ is the spin degeneracy and $g_v$ is the valley degeneracy.
The quantum mobility is given by $\mu_{q} = e \tau_{q}/m^{\star}$, where the quantum (single particle) scattering rate reads
\begin{align}
	\frac{1}{\tau_{q}} =\frac{2m^{\star}}{\pi \hbar^3}\int \limits_0^{2k_F}\frac{dq}{\sqrt{4k_F^2 - q^2}}\, \ev{\abs{U(q)}^2}. \label{eq:1_tauq}
\end{align}
The quantum scattering rate can be indirectly measured through the Dingle temperature $T_D = \hbar/2\pi \tau_q$ which characterizes the damping of Shubnikov-de Haas (SdH) oscillations. 
In general, $\tau_q < \tau$ since the transport mobility ignores all forward scattering. (For strictly zero range s-wave scattering the two scattering times are trivially equal, but for long-range disorder scattering the two could vary by orders of magnitude.)
For charged impurity scattering, the potential correlator averaged over impurity positions is given by 
\begin{align}\label{eq:uq2}
    \ev{\abs{U(q)}^2} = \int \limits_{-\infty}^{+\infty} dz \; N(z) U_1^2(q,z)\,,
\end{align}
where $N(z)$ is the 3D concentration of impurities at a distance $z$ from the center of the 2DEG, and $U_1(q,z)$ is the screened Coulomb potential for an impurity located at $z$:
\begin{align}\label{eq:u_i1_mu}
    U_1(q,z) = \frac{2\pi e^2}{\kappa q \epsilon(q) }
    \int \limits_{-\infty}^{+\infty} dz' \;\abs{\psi(z')}^2 e^{-q\abs{z- z'}}\,,
\end{align}
where $\psi(z)$ is the confinement wave function along $z$ direction and the dielectric function is given by
\begin{align}\label{eq:df}
    \epsilon(q) = 1 + (q_{\rm TF}/q) F_c (qw) [1-G(q)] \,.
\end{align}
Here, $q_{TF} = g/a_B$ is the Thomas-Fermi screening wave vector, $a_B = \kappa \hbar^2 / m^{\star} e^2$ is the effective Bohr radius, $\kappa = 11.9$ is the dielectric constant of Si, and $w$ is the QW width. 
$G(q) = q/(g \sqrt{q^2 + k_F^2})$ is the local field correction using the Hubbard approximation~\cite{jonson:1976,Gold:1994} which reflects the suppressed screening at $q\gtrsim k_F$ (with $G=0$ being the RPA, which overestimates short wavelength screening). 
The form factor $F_c(qw)$ is given by
\begin{equation}\label{eq:fcq}
    F_c(qw) = \iint \limits_{-\infty}^{+\infty} dz dz' \abs{\psi(z)}^2  \abs{\psi(z')}^2 \exp(-q\abs{z-z'})\,.
\end{equation}
In Eq.~\eqref{eq:u_i1_mu} we ignore the screening from the top gate. 
To include the gate screening, we should add the Coulomb potential of the image charges and replace $e^{-q\abs{z- z'}}$ by $\qty(e^{-q\abs{z- z'}} - e^{-q\abs{2d_g - z - z'}})$, where $d_g = d + d_o + w/2$ is the distance from the gate to the center of the 2DEG.

\section{delta layer 2DEG approximation}
\label{sec:delta}
For simplicity, assuming that the gate is far away $d_g \gg d$ and the wavefunction perpendicular to the quantum well has a delta function profile $\abs{\psi(z)}^2 = \delta(z)$, the form factors reduce to identities and $U_1(q,z) = \frac{2\pi e^2}{\kappa q \epsilon(q) } e^{-q\abs{z}}$.
In this section, we disregard the local field correction, thus making the dielectric function the RPA form, $\epsilon(q) = 1 + q_{TF}/q$. 
We will address the effect of form factors and local field correction in Section~\ref{sec:inf}, and the effect of gate screening in Section~\ref{sec:gate}.
We use a 2-impurity model to calculate the scattering rate following Refs.~\cite{Hwang:2013,Hwang:2014}.
That is, we assume that there are two delta layers of impurities. One layer is located at a distance $d$ from the 2DEG, which represents the remote charged impurities in the capping layer. 
The other layer is at $z=0$, which represents oxygen-related background charges in the QW~\cite{Petta:2015} or an effective description of other short-range scattering sources such as the interface roughness~\cite{DC_Tsui:2012,Hwang:2014_short_range}. 
The impurity distribution is given by
\begin{align}\label{eq:model1}
    N(z) = n_r\delta(z-d) + n_2 \delta(z).
\end{align}
The corresponding transport scattering rate $\tau^{-1} = \tau_{\mathrm{RI}}^{-1} + \tau_{\mathrm{S}}^{-1}$ reads~\cite{Hwang:2013a_valley,Hwang:2013,Hwang:2014}
\begin{gather}\label{eq:rate1_nint}
    \frac{1}{\tau_{\mathrm{RI}}} = n_r\frac{2\pi\hbar}{m^{\star}} \qty(\frac{2}{g})^2 f\qty(2k_Fd, \frac{q_{TF}}{2k_F}), \\
    \frac{1}{\tau_{\mathrm{S}}} = n_2\frac{2\pi\hbar}{m^{\star}} \qty(\frac{2}{g})^2 f\qty(0,\frac{q_{TF}}{2k_F}). \label{eq:rate2_nint}
\end{gather}
The quantum scattering rate $\tau_q^{-1} = \tau_{q\mathrm{RI}}^{-1} + \tau_{q\mathrm{S}}^{-1}$ reads
\begin{align}\label{eq:rate1q_nint}
    \frac{1}{\tau_{q\mathrm{RI}}} = n_r\frac{2\pi\hbar}{m^{\star}} \qty(\frac{2}{g})^2 f_{q}\qty(2k_Fd, \frac{q_{TF}}{2k_F}), \\
    \frac{1}{\tau_{q\mathrm{S}}} = n_2\frac{2\pi\hbar}{m^{\star}} \qty(\frac{2}{g})^2 f_{q}\qty(0,\frac{q_{TF}}{2k_F}), \label{eq:rate2q_nint}
\end{align}
and the dimensionless functions $f(a,s)$ and $f_{q}(a,s)$ are defined as
\begin{align}
    f(a,s) = \int_{0}^{1} s^2\frac{2x^2 e^{-2ax}\,dx}{\sqrt{1-x^2}(x+s)^2}, \\
    f_{q}(a,s) = \int_{0}^{1} s^2\frac{e^{-2ax}\,dx}{\sqrt{1-x^2}(x+s)^2}.
\end{align}
%First, let us discuss the strong screening limit $q_{TF}/2k_F \gg 1$. 
In the density range
\begin{align}\label{eq:density_RI}
    \frac{g}{16 \pi d^2} \ll n_e \ll \frac{g^3}{16 \pi a_B^2},
\end{align}
the limits $2k_F d \gg 1$ and $q_{TF}/2k_F \gg 1$ are satisfied and we have $f(a,s) = (2a^3)^{-1}$ and $f_{q}(a,s) = (2a)^{-1}$~\cite{RI_integral} which lead to the well-known results for RI scattering~\cite{Price:1981,Gold:1988,Monroe:1993,Dmitriev:2012}
\begin{gather}\label{eq:rate1}
    \frac{1}{\tau_{\mathrm{RI}}} = n_r \qty(\frac{2}{g})^2 \frac{\pi\hbar}{8m^{\star}(k_Fd)^3} , \\
    \frac{1}{\tau_{q\mathrm{RI}}} = n_r \qty(\frac{2}{g})^2 \frac{\pi\hbar}{2 m^{\star}k_Fd}. \label{eq:rate1q}
\end{gather}
and the corresponding ratio is given by
\begin{align}\label{eq:ratio_RI}
    \qty(\frac{\tau}{\tau_q})_{\mathrm{RI}} = (2k_F d)^2 \gg 1.
\end{align}
The large difference between $\tau$ and $\tau_q$ reflects the diffusive character of the electron dynamics on the Fermi surface scattered from remote impurities~\cite{Stern:1985}. 
The physical meaning of $\tau/\tau_q$ is the typical number of small-momentum forward scattering events needed to change the direction of momentum by an angle of order $\pi$~\cite{Dmitriev:2012}.
Next, we calculate the short-range scattering rate $\tau_{\mathrm{S}}^{-1}$ and $\tau_{q\mathrm{S}}^{-1}$ contributed from a delta layer of impurities at $z=0$.
The related integrals $f(0,s)$ and $f_{q}(0,s)$ at $s>1$ can be calculated analytically:
\begin{gather}
    f(0,s) = s^2\pi + \frac{2 s^3}{1-s^2}+\frac{2 s^3\left(2-s^2\right) \sec ^{-1}(s)}{\left(s^2-1\right)^{3/2}}, \\
    f_{q}(0,s) = \frac{s}{1-s^2}+\frac{s^3 \, \mathrm{sec}^{-1}(s)}{ \left(s^2-1\right)^{3/2}}.
\end{gather}
In the density range $n_e \ll g^3/16 \pi a_B^2$, the limit $s=q_{TF}/2k_F \gg 1$ is satisfied and $f(0,s)=f_{q}(0,s) = \frac{\pi}{2}$, so that $\tau_{\mathrm{S}}^{-1}$ and $\tau_{q\mathrm{S}}^{-1}$ are given by~\cite{Hwang:2013,Hwang:2013a_valley}
\begin{align}\label{eq:rate2}
    \frac{1}{\tau_{\mathrm{S}}} = \frac{1}{\tau_{q\mathrm{S}}} = n_2\frac{\pi^2 \hbar}{m^{\star}} \qty(\frac{2}{g})^2,
\end{align}
which is independent of the electron concentration.
The ratio $\tau/\tau_q = 1$ reflects that short-range impurities are efficient in backscattering the electrons -- this is the situation in regular 3D metals where $\tau$ and $\tau_q$ are basically the same.
%Next, we examine the weak screening limit $q_{TF}/2k_F \ll 1$, where the density is high $n_e \gg \frac{g^3}{16 \pi a_B^2}$.

\begin{figure*}[t]
    \centering
    \includegraphics[width = 0.8\linewidth]{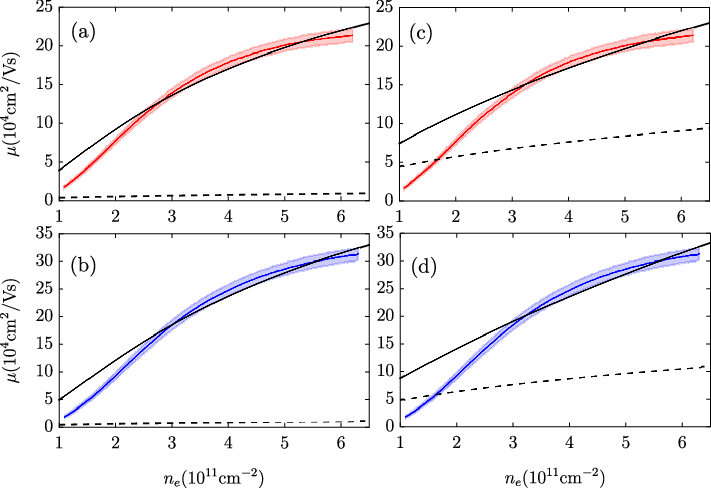}%mu_low_ne0.eps
    \caption{Mobility ($\mu$) versus electron density ($n_e$) for quantum wells with widths of 5 nm (red) and 7 nm (blue). The experimental data are depicted by red and blue curves, with corresponding error bars shown as shaded areas. The solid black curves represent the theoretical transport mobility, while the dashed black curves denote the quantum mobility, both calculated using the delta-layer approximation for the 2DEG without local field correction. Subfigures (a) and (b) incorporate remote and short-range impurity scattering [cf. Eqs.~\eqref{eq:rate1_nint}, \eqref{eq:rate2_nint}, and \eqref{eq:rate1q_nint}], whereas (c) and (d) consider background and short-range impurity scattering [cf. Eqs.~\eqref{eq:rate2_nint}, \eqref{eq:rate3_nint}, and \eqref{eq:rate3q_nint}]. Fitting parameters: (a) $n_r = 1.9 \times 10^{12}$ cm$^{-2}$ and $n_2= 2.8 \times10^{9}$ cm$^{-2}$; (b) $n_r = 1.7 \times 10^{12}$ cm$^{-2}$ and $n_2= 1.8 \times10^{9}$ cm$^{-2}$; (c) $n_2= 2.1 \times10^{9}$ cm$^{-2}$ and $N_1= 11\times10^{15}$ cm$^{-3}$; (d) $n_2= 1.4\times10^{9}$ cm$^{-2}$ and $N_1= 12\times10^{15}$ cm$^{-3}$.}
    \label{fig:mobility_nint_delta}
\end{figure*}

In the other 2-impurity model, we replace the delta-layer remote impurities by uniform background charged impurities outside the well with a 3D concentration $N_1$.
The impurity distribution is given by
\begin{align}\label{eq:model2}
    N(z) =  N_1 \Theta(\abs{z} - w/2) + n_2 \delta(z),
\end{align}
The corresponding scattering rates due to uniform background charged impurities are given by
\begin{align}\label{eq:rate3_nint}
    \frac{1}{\tau_{\mathrm{BI}}} = N_1\frac{2 \pi\hbar}{m^{\star}} \qty(\frac{2}{g})^2 \int\limits_{\abs{z} > w/2}  f\qty(2k_F z, \frac{q_{TF}}{2k_F}) dz, \\
    \frac{1}{\tau_{q\mathrm{BI}}} = N_1\frac{2\pi\hbar}{m^{\star}} \qty(\frac{2}{g})^2 \int\limits_{\abs{z} > w/2} f_{q}\qty(2k_F z, \frac{q_{TF}}{2k_F}) dz. \label{eq:rate3q_nint}
\end{align}
In the electron density range
\begin{align}\label{eq:density_BI_large}
    \frac{g}{16 \pi w^2} \ll n_e \ll \frac{g^3}{16 \pi a_B^2},
\end{align}
the limits $2k_Fw \gg 1$ and $q_{TF}/2k_F \gg 1$ are satisfied and we get the analytical expressions of the BI scattering rates
\begin{align}\label{eq:rate3}
    \frac{1}{\tau_{\mathrm{BI}}} = N_1 w \qty(\frac{2}{g})^2 \frac{\pi\hbar }{2 m^{\star}(k_F w)^3} , \\
    \frac{1}{\tau_{q\mathrm{BI}}} = N_1 w \qty(\frac{2}{g})^2 \frac{\pi\hbar \ln(2d_s/w)}{m^{\star}k_F w}, \label{eq:rate3q}
\end{align}
where $d_s$ is the cutoff distance of the BI distribution so that impurities at $\abs{z}>d_s$ can be ignored. 
Otherwise, $\tau_{q\mathrm{BI}}^{-1}$ diverges logarithmically due to the contribution of small-momentum scattering from the long distance. 
We choose $d_s = d_g/2$ as half the distance from the gate to the center of the 2DEG, considering that charged impurities close to the gate should be effectively screened~\cite{Sammon:2018}. 
We will discuss this choice of $d_s$ and the gate screening effect in more detail in Section~\ref{sec:gate}.

In the density range described by Eqs.~\eqref{eq:density_RI} and \eqref{eq:density_BI_large}, we see from Eqs.~\eqref{eq:rate1} and \eqref{eq:rate3} that $\tau_{\mathrm{RI}}^{-1}$ and $\tau_{\mathrm{BI}}^{-1}$ are proportional to $k_F^{-3}$, which leads to a mobility proportional to $n_e^{3/2}$.
On the other hand, $\tau_{\mathrm{S}}^{-1}$ in Eq.~\eqref{eq:rate2} does not depend on $k_F$, so it results in a constant mobility that does not depend on $n_e$.
As a result, the total mobility described by the 2-impurity model $\tau^{-1} = \tau_{\mathrm{RI}}^{-1} + \tau_{\mathrm{S}}^{-1}$ or $\tau^{-1} = \tau_{\mathrm{BI}}^{-1} + \tau_{\mathrm{S}}^{-1}$ should increase as $n_e^{3/2}$ at low density and eventually saturate at high density.
We can compare the density range $n_e = 1-6 \times 10^{11}$ cm$^{-2}$ in the high mobility structures reported in Refs.~\cite{Wuetz:2023,Esposti:2023} with the density ranges written in Eqs.~\eqref{eq:density_RI} and \eqref{eq:density_BI_large}.
We use the spin/valley degeneracy $g_s = 2$ and $g_v = 2$ for the high mobility structures reported in Refs.~\cite{Wuetz:2023,Esposti:2023}.
In the density range $n_e = 1-6 \times 10^{11}$ cm$^{-2}$ the Fermi energy calculated using $g=4$ is $E_F = 0.6-3.8$ meV which is much higher than the valley splitting $0.2$ meV reported in Refs.~\cite{Wuetz:2023,Esposti:2023}, so our choice of $g=4$ is correct.
In Section~\ref{sec:valley}, we will discuss the effects of quantum degeneracy on the transport properties in more detail.
Using $g=4$, we find that the experimental density range is well within the range described in Eq.~\eqref{eq:density_RI} where the upper bound is $g^3/16\pi a_B^2 \approx 10^{13}$ cm$^{-2}$ and the lower bound is $g/16\pi d^2 \approx 10^{10}$ cm$^{-2}$.
Therefore, the asymptotic expressions for RI scattering Eqs.~\eqref{eq:rate1} and \eqref{eq:rate1q} should be a good approximation. 
On the other hand, the lower bound in Eq.~\eqref{eq:density_BI_large} is $g/16\pi w^2 \approx 2.4 \times 10^{11}$ cm$^{-2}$, which means that the asymptotic expressions for BI scattering rates Eqs.~\eqref{eq:rate3} and \eqref{eq:rate3q} are good only when $n_e \gg 2.4 \times 10^{11}$ cm$^{-2}$.
At low densities $n_e \ll 2.4 \times 10^{11}$ cm$^{-2}$, we have 
\begin{align}\label{eq:rate3_low_density}
    \frac{1}{\tau_{\mathrm{BI}}} = \frac{1}{\tau_{q\mathrm{BI}}} = N_1(2d_s - w)\frac{\pi^2 \hbar}{m^{\star}} \qty(\frac{2}{g})^2,
\end{align}
and the background impurity scattering behaves similarly to short-range impurity scattering with an effective 2D impurity concentration $N_1(2d_s - w)$.
Hence, as the density is reduced, $\tau_{\mathrm{BI}}^{-1}$ and $\tau_{q\mathrm{BI}}^{-1}$ transition from a long-range behavior, as expressed in Eqs.~(\ref{eq:rate3}-\ref{eq:rate3q}), to a short-range behavior, as expressed in Eq.~\eqref{eq:rate3_low_density}.
The crossover electron density is at $n_e = 2.4 \times 10^{11}$ cm$^{-2}$.

Fig.~\ref{fig:mobility_nint_delta} shows the results of the best-fit mobility curves (solid black) and the corresponding prediction of quantum mobility (dashed black) by performing the numerical integrals of Eqs.~\eqref{eq:rate1_nint}, \eqref{eq:rate2_nint} and \eqref{eq:rate3_nint}.
The experimental data from Refs.~\cite{Wuetz:2023,Esposti:2023}, represented by blue and red curves with shaded areas indicating error bars, are also included for comparison.
The results of the combination of remote and short-range impurity scattering are shown in Figs.~\ref{fig:mobility_nint_delta} (a) and (b), while the results of the combination of background and short-range impurity scattering are shown in Figs.~\ref{fig:mobility_nint_delta} (c) and (d).
The fitting impurity densities are $n_r \approx 2\times 10^{12}$ cm$^{-2}$, $n_2 \approx 1.4-2.8\times 10^{9}$ cm$^{-2}$ and $N_1 \approx 10 \times 10^{15}$ cm$^{-3}$, in reasonable agreement with the numbers reported in Refs.~\cite{Stern:1992,Tanner:2007,Gold_2010,Petta:2015,Hwang:2013a_valley,Laroche:2015,Ahn:2022}.
In both 2-impurity models, short-range impurity scattering is the limiting factor for transport mobility at high densities $n_e = 4-6 \times 10^{11}$ cm$^{-2}$.
The dominant scattering mechanism at lower densities is RI or BI, depending on the choice of the impurity model. 
We see that the first 2-impurity model ($\tau^{-1} = \tau_{\mathrm{RI}}^{-1} + \tau_{\mathrm{S}}^{-1}$) fits the mobility data at intermediate to high electron density $n_e = 2.5-6 \times 10^{11}$ cm$^{-2}$ within the experimental uncertainty.
At the highest density $n_e = 6 \times 10^{11}$ cm$^{-2}$ for both quantum wells, the predicted quantum mobility is around $1.0 \times 10^4$ cm$^2$/Vs, while the measurements of SdH oscillations give a larger value $(3.0 \pm 0.5) \times 10^4$ cm$^2$/Vs~\cite{private_comm}.
On the other hand, the other 2-impurity model ($\tau^{-1} = \tau_{\mathrm{BI}}^{-1} + \tau_{\mathrm{S}}^{-1}$) fits the experimental mobility data in a narrower density range $n_e =3-6 \times 10^{11}$ cm$^{-2}$, with the predicted quantum mobility around $10 \times 10^4$ cm$^2$/Vs at the highest density $n_e = 6 \times 10^{11}$ cm$^{-2}$, which is a factor of 3 larger than the experimental value~\cite{private_comm}.
%This difference in the predicted quantum mobility is due to the typical momentum transfer being much smaller in RI scattering than in BI scattering, leading to a significantly larger ratio of $(\mu/\mu_q)_{\mathrm{RI}}$ compared to $(\mu/\mu_q)_{\mathrm{BI}}$~\cite{Stern:1985}.
Since the typical momentum transfer of RI scattering is much smaller than that of BI scattering, the ratio $(\mu/\mu_q)_{\mathrm{RI}} \gg (\mu/\mu_q)_{\mathrm{BI}}$~\cite{Stern:1985}, which explains a factor of 10 difference in the predicted quantum mobility shown in Fig.~\ref{fig:mobility_nint_delta}.
%In Section~\ref{sec:gate}, we find that a better agreement on quantum mobility can be reached by including gate screening effects in remote impurity scattering calculations, because the gate screens small-momentum scattering and increases quantum mobility. 
In Section~\ref{sec:gate}, we will explore how including gate screening effects in RI scattering calculations can improve the agreement with experimental quantum mobility data. 
However, it is important to note that the experimental accuracy for quantum mobility data is generally less reliable than for transport mobility, largely due to the indirect method of extracting quantum mobility through the Dingle temperature~\cite{Stern:1985}. 
%However, we should emphasize that the experimental accuracy for the quantum mobility data in general is much less reliable than that for the transport mobility, because the extraction of the quantum mobility is rather indirect through the Dingle temperature~\cite{Stern:1985}.
%The semi-quantitative agreement between our theory and the experimental data should be satisfactory.
Given these considerations, a semi-quantitative agreement between our theoretical predictions and experimental data should be regarded as satisfactory.
At low densities $n_e <2 \times 10^{11}$ cm$^{-2}$, the theoretical mobility curves deviate and become higher than the experimental data.
We should emphasize that this deviation at low densities does not arise from any new scattering mechanism ignored in the theory.
The dominant scattering mechanism at low densities $n_e < 4 \times 10^{11}$ cm$^{-2}$ is RI if we use the 2-impurity model Eq.~\eqref{eq:model1} [or BI if we use the 2-impurity model Eq.~\eqref{eq:model2}].
Rather, this deviation indicates the systematic failure of the Boltzmann-Born theory at low densities, a topic we will address in more detail in Section~\ref{sec:mit}, particularly in the context of the metal-insulator transition.
%Such a deviation indicates the breakdown of the Boltzmann-Born theory at low densities, and we will discuss the corresponding metal-insulator transition in more detail in Section~\ref{sec:mit}.

\begin{figure*}[t]
    \centering
    \includegraphics[width = \linewidth]{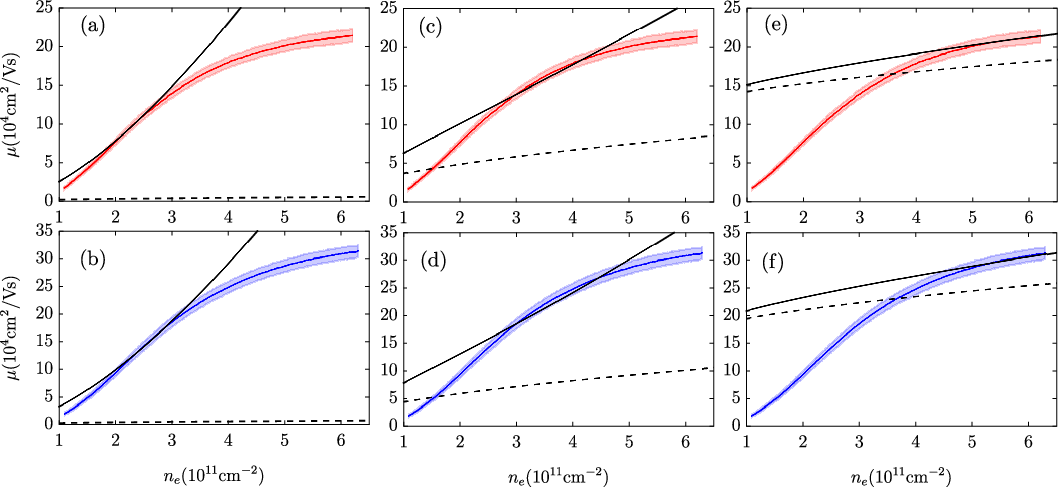}%mu_low_ne0.eps
    \caption{Mobility ($\mu$) versus electron density ($n_e$) for quantum wells with widths of 5 nm (red) and 7 nm (blue). The experimental data are depicted by red and blue curves, with corresponding error bars shown as shaded areas. The solid black curves represent the theoretical transport mobility, while the dashed black curves denote the quantum mobility, both calculated using the infinite potential well approximation without the local field correction. (a) and (b) assume only RI near the capping layer $n_r \neq 0$. (c) and (d) assume only BI outside the quantum well $N_1 \neq 0$ and no BI inside $N_2=0$, while (e) and (f) assume $N_1 = 0$ and $N_2 \neq 0$.  Impurity concentrations and roughness parameters for fitting: (a) $n_r = 3.1 \times 10^{12}$ cm$^{-2}$; (b) $n_r = 2.6 \times 10^{12}$ cm$^{-2}$; (c) $N_1=14\times10^{15}$ cm$^{-3}$ and $N_2= 0$; (d) $N_1=12.6\times10^{15}$ cm$^{-3}$ and $N_2=0$; (e) $N_1=0$ and $N_2= 10.3\times10^{15}$ cm$^{-3}$; (f) $N_1=0$ and $N_2=5.5\times10^{15}$ cm$^{-3}$.}
    \label{fig:mobility_infinite_noG_individual}
\end{figure*}
\begin{figure*}[t]
    \centering
    \includegraphics[width = 0.8\linewidth]{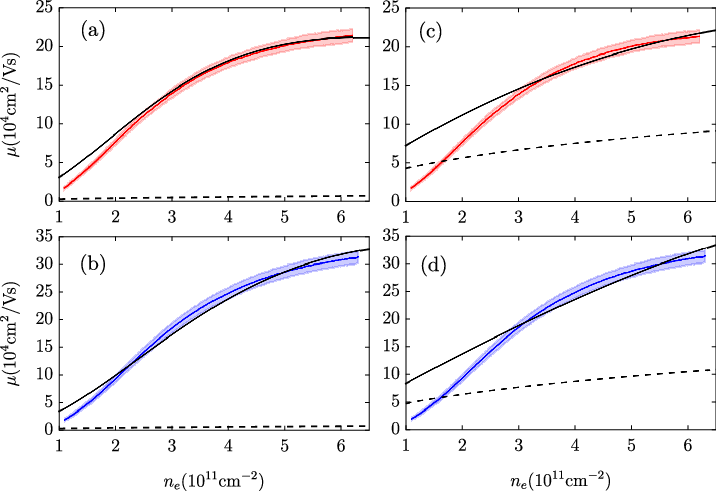}%mu_low_ne0.eps
    \caption{Mobility ($\mu$) versus electron density ($n_e$) for quantum wells with widths of 5 nm (red) and 7 nm (blue). The experimental data are depicted by red and blue curves, with corresponding error bars shown as shaded areas. The solid black curves represent the theoretical transport mobility, while the dashed black curves denote the quantum mobility, both calculated using the infinite potential well approximation without the local field correction. Subfigures (a) and (b) incorporate remote impurity and interface roughness scattering, whereas (c) and (d) consider background impurity and interface roughness scattering. Impurity concentrations and roughness parameters for fitting: (a) $n_r = 2.5 \times 10^{12}$ cm$^{-2}$, $\Delta = 4.5$ {\AA}, and $\Lambda = 27$ {\AA}; (b) $n_r = 2.5 \times 10^{12}$ cm$^{-2}$, $\Delta = 8$ {\AA}, and $\Lambda = 28$ {\AA}; (c) $N_1= 12\times10^{15}$ cm$^{-3}$, $N_2=0$, $\Delta = 5$ {\AA}, and $\Lambda = 16$ {\AA}; (d) $N_1= 12\times10^{15}$ cm$^{-3}$, $N_2=0$, $\Delta = 7$ {\AA}, and $\Lambda = 20$ {\AA}.}
    \label{fig:mobility_infinite_noG_all}
\end{figure*}

\begin{figure*}[t]
    \centering
    \includegraphics[width = \linewidth]{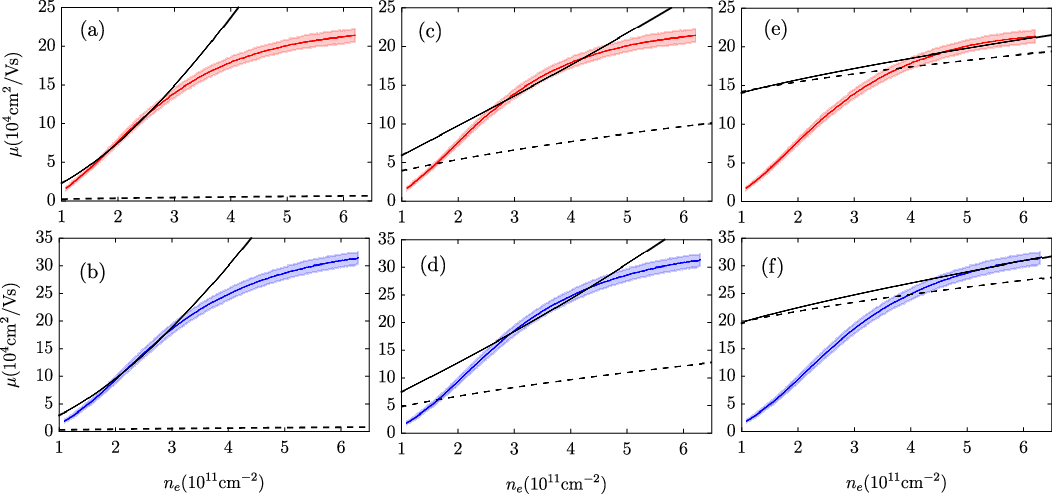}%mu_low_ne0.eps
    \caption{Mobility ($\mu$) versus electron density ($n_e$) for quantum wells with widths of 5 nm (red) and 7 nm (blue). The experimental data are depicted by red and blue curves, with corresponding error bars shown as shaded areas. The solid black curves represent the theoretical transport mobility, while the dashed black curves denote the quantum mobility, both calculated using the infinite potential well approximation with the local field correction. (a) and (b) assume only RI near the capping layer $n_r \neq 0$. (c) and (d) assume only BI outside the quantum well $N_1 \neq 0$ and no BI inside $N_2=0$, while (e) and (f) assume $N_1 = 0$ and $N_2 \neq 0$. To fit the data, the impurity concentrations and roughness parameters are chosen as (a) $n_r = 2.5 \times 10^{12}$ cm$^{-2}$; (b) $n_r = 2.1 \times 10^{12}$ cm$^{-2}$; (c)  $N_1=9.7\times10^{15}$ cm$^{-3}$ and $N_2=0$;  (d) $N_1=8.7\times10^{15}$ cm$^{-3}$ and $N_2=0$; (e) $N_1=0$ and $N_2= 7.2\times10^{15}$ cm$^{-3}$; (f) $N_1=0$ and $N_2=3.8\times10^{15}$ cm$^{-3}$.}
    \label{fig:mobility_infinite_G_individual}
\end{figure*}
\begin{figure*}[t]
    \centering
    \includegraphics[width = 0.8\linewidth]{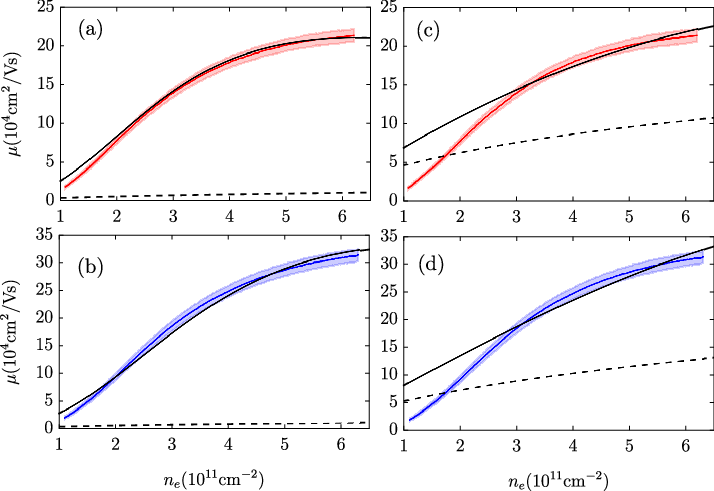}%mu_low_ne0.eps
    \caption{Mobility ($\mu$) versus electron density ($n_e$) for quantum wells with widths of 5 nm (red) and 7 nm (blue). The experimental data are depicted by red and blue curves, with corresponding error bars shown as shaded areas. The solid black curves represent the theoretical transport mobility, while the dashed black curves denote the quantum mobility, both calculated using the infinite potential well approximation with the local field correction. Subfigures (a) and (b) incorporate remote impurity and interface roughness scattering, whereas (c) and (d) consider background impurity and interface roughness scattering. Fitting parameters: (a) $n_r = 2.0 \times 10^{12}$ cm$^{-2}$, $\Delta = 3.5$ {\AA}, and $\Lambda = 30$ {\AA}; (b) $n_r = 2.0 \times 10^{12}$ cm$^{-2}$, $\Delta = 6.5$ {\AA}, and $\Lambda = 30$ {\AA}; (c) $N_1= 8.1\times10^{15}$ cm$^{-3}$, $N_2=0$, $\Delta = 3$ {\AA}, and $\Lambda = 22$ {\AA}; (d) $N_1= 8.1\times10^{15}$ cm$^{-3}$, $N_2=0$, $\Delta = 6$ {\AA}, and $\Lambda = 22$ {\AA}.}
    \label{fig:mobility_infinite_G_all}
\end{figure*}

\section{Infinite potential well}
\label{sec:inf}
In this section, we use an infinite potential well approximation in order to address the quantum well confinement better, where the 2DEG wave function is given by $\psi(z) = \sqrt{2/w} \cos(\pi z/w) \Theta(w/2 - \abs{z})$~\cite{Gold:2010,Hwang:2013}. 
The justification of the infinite potential well approximation will be discussed at the end of this section.
Substituting this wavefunction into Eq.~\eqref{eq:u_i1_mu}, we obtain the single-impurity Coulomb potential
\begin{align}\label{eq:U1_inf}
    {U}_{1}(q,z) = \frac{2\pi e^2}{\kappa q \epsilon(q)}
    \begin{cases}
        F_0(qw) e^{-q\abs{z}}, & \abs{z} > w/2\,, \\
        G_0(q,z), & \abs{z} < w/2\,,
    \end{cases}
\end{align}
where 
\begin{align}
    F_0(x) = \frac{8 \pi^2}{x(x^2 + 4\pi^2)} \sinh(x/2)\, ,
\end{align}
and
\begin{align}
    G_0(q,z) &=\frac{8\pi^2 (1 - e^{-qw/2} \cosh{qz}) }{qw(4\pi^2 + q^2 w^2)} \nonumber \\
    &+ \frac{2q w [1+ \cos(2\pi z /w)]}{4\pi^2 + q^2 w^2} \,.
\end{align}
Both $F_0(qw)$ and $G_0(q,z)$ monotonically decrease as a function of $q$ and weaken the scattering potential at large momentum $q$.
The screening form factor $F_c(qw)$ in Eq.~\eqref{eq:fcq} reads
\begin{align}
    F_c(x) = \frac{20 \pi^2 x^3 + 3 x^5 - 32 \pi^4 (1-e^{-x} - x)}{x^2(4\pi^2 + x^2)^2}.
\end{align}
This form factor enters the dielectric function $\epsilon(q)$ and weakens the screening at large momentum $q$.

%We tried two different 2-impurity models as described in Section.~\ref{sec:delta} to interpret the experimental data. 
First, we calculate the charged impurity scattering rates separately and see how well they can interpret the data individually.
The 3D impurity concentration is given by 
\begin{align}
    N(z) &= n_r \delta(z-d-w/2) +\nonumber \\
    &+ N_1 \Theta(\abs{z} - w/2) + N_2 \Theta(w/2 - \abs{z}), \label{eq:BG_Nz}
\end{align}
where $n_r$ is the RI concentration in the capping layer, $N_1$ ($N_2$) is the BI concentration outside (inside) the quantum well.
The theoretical results of mobility and quantum mobility are shown in Fig.~\ref{fig:mobility_infinite_noG_individual}, based on the assumption that only one term in Eq.~\eqref{eq:BG_Nz} is nonzero.
Figs.~\ref{fig:mobility_infinite_noG_individual} (a) and (b) show the results of $n_r \neq 0$ and $N_1 = N_2 = 0$.
This RI-dominant mobility can fit the data at intermediate densities $n_e =2-3 \times 10^{11}$ cm$^{-2}$ within the margin of experimental uncertainty. 
The predicted quantum mobility is around $1.0\times 10^{4}$ cm$^{2}$/Vs at the highest density $n_e = 6 \times 10^{11}$ cm$^{-2}$.
Figs.~\ref{fig:mobility_infinite_noG_individual} (c) and (d) show the results of $N_1 \neq 0$ and $N_2 =n_r = 0$, which fit the data at intermediate densities $n_e =2.5-5 \times 10^{11}$ cm$^{-2}$. 
The predicted quantum mobility is around $10 \times 10^{4}$ cm$^{2}$/Vs at the highest density $n_e = 6 \times 10^{11}$ cm$^{-2}$.
Figs.~\ref{fig:mobility_infinite_noG_individual} (e) and (f) show the results of $N_2 \neq 0$ and $N_1 =n_r = 0$, which fit the data at high densities $n_e=4-6 \times 10^{11}$ cm$^{-2}$. 
The corresponding ratio $\tau/\tau_q \approx 1$ is expected, as the background impurities inside the QW should behave similarly to the short-range impurities because of strong screening by the electrons.
In general, switching the dominant impurity distribution from RI ($n_r\neq 0$) to BI ($N_1 \neq 0$) and finally to a different BI distribution ($N_2 \neq 0$) results in a progressive flattening of the mobility $\mu(n_e)$ and a decrease in the ratio $\mu/\mu_q$ towards unity. 
%This trend indicates a crossover from remote impurity scattering to short-range impurity scattering.
%In general, by switching the dominant impurity distribution from $n_r\neq 0$ to $N_1 \neq 0$ and finally to $N_2 \neq 0$, the mobility curve $\mu(n_e)$ becomes flatter and the ratio $\mu/\mu_q$ becomes smaller and approaches unity, which shows a crossover from remote impurity scattering to short-range impurity scattering.
%On the other hand, the corresponding quantum mobility at the highest density $n_e = 6 \times 10^{11}$ cm$^{-2}$ is $\mu_q = 13 \times 10^{4}$ cm$^{2}$/Vs for a 5 nm QW and $\mu_q = 18 \times 10^{4}$ cm$^{2}$/Vs for a 7 nm QW , which are 2 - 3 times higher than the quantum mobility calculated due to only $N_1\neq 0$.

%Therefore, $N_2$ is probably not the dominated scattering mechanism which limits the mobility at intermediate to high density.

%Power law increases to $\alpha \approx 2$ by a local field correction $G(q) = q/2\sqrt{q^2 + k_F^2}$~\cite{Dolgopolov:2003,Gold:2010,Khanh:2013,Laroche:2015}. Double the Hubbard form~\cite{Dolgopolov:2003,Khanh:2013}.

%Low-density theory should be described by something else like a percolation theory~\cite{Tracy:2009}. 

In order to better fit the mobility data, we use interface roughness as an additional scattering mechanism that limits mobility at high density~\cite{Gold:1988,Feenstra:1995,Penner:1998,Hwang:2004,Tracy:2009,Gold:2010,Hwang:2014_short_range}.
Using the correlator of local well width variations $\ev{\Delta(\vb{r}) \Delta(\vb{r}')} = \Delta^2 \exp(-\abs{\vb{r} - \vb{r}'}^2/\Lambda^2)$, where $\Delta$ is the typical roughness height and $\Lambda$ is the roughness lateral size, the scattering potential due to interface roughness is given by 
\begin{align}\label{eq:usq2}
    \ev{\abs{U_{\rm IR}(q)}^2} = \frac{\pi}{\epsilon^2(q)} \Delta^2 \Lambda^2 \qty(\pdv{E}{w})^2 e^{-q^2\Lambda^2/4}\,,
\end{align}
where $E$ is the ground state energy of the potential well along $z$ direction. 
For an infinite potential well, the ground state energy $E = E_0 = \hbar^2 \pi^2/2m^\star_z w^2$, and $\partial E/\partial w = 2E_0/w$. 
Here $m_z^{\star} = 0.92 m_0$ is the longitudinal effective mass of an electron in Si.
Using Eqs.~\eqref{eq:usq2}, \eqref{eq:1_tau}, and \eqref{eq:1_tauq}, we obtain the transport relaxation rate $\tau_{\mathrm{IR}}^{-1}$ and quantum rate $\tau_{q,\mathrm{IR}}^{-1}$ due to interface roughness scattering.
At low electron densities $n_e \ll g/4\pi \Lambda^2$ where $k_F \Lambda \ll 1$ and $q_{TF}/2k_F \gg 1$ are satisfied, we obtain
\begin{align}\label{eq:rate_IR_low_density}
    \tau_{\mathrm{IR}}^{-1} = \frac{1}{\tau_{0}}\frac{3\pi}{8}   \qty(\frac{2k_F}{q_{TF}})^2, \\
    \tau_{q\mathrm{IR}}^{-1} = \frac{1}{\tau_{0}} \frac{\pi}{4}   \qty(\frac{2k_F}{q_{TF}})^2,\label{eq:rate_qIR_low_density}
\end{align}
where $\tau_{0}^{-1}$ is given by
\begin{align}
    \tau_{0}^{-1} = \frac{ 2 \pi^4 \hbar}{m^{\star}} \qty(\frac{m^{\star}}{m^{\star}_z})^2 \frac{\Delta^2 \Lambda^2}{w^6}.
\end{align}
We find that both $\mu_{\mathrm{IR}}$ and $\mu_{q\mathrm{IR}}$ are proportional to $n_e^{-1}$ at low electron densities $n_e \ll g/4\pi \Lambda^2$.
In the high-density range $g/4\pi \Lambda^2 \ll n_e \ll g^3 /16\pi a_B^2$, we obtain
\begin{align}\label{eq:rate_IR_high_density}
    \tau_{\mathrm{IR}}^{-1} = \frac{1}{\tau_{0}} \frac{3\sqrt{\pi}}{4}   \qty(\frac{2k_F}{q_{TF}})^2 (k_F \Lambda)^{-5}, \\
    \tau_{q\mathrm{IR}}^{-1} = \frac{1}{\tau_{0}} \frac{\sqrt{\pi}}{4}   \qty(\frac{2k_F}{q_{TF}})^2 (k_F \Lambda)^{-3},\label{eq:rate_qIR_high_density}
\end{align}
where we find $\mu_{\mathrm{IR}} \propto n_e^{3/2}$ and $\mu_{q\mathrm{IR}} \propto n_e^{1/2}$~\cite{IR_integral}, similar to the long-range remote impurity scattering results [cf. Eqs.~\eqref{eq:rate1} and \eqref{eq:rate1q}].
The crossover of the $\mu_{\mathrm{IR}} \propto n_e^{-1}$ to the $\mu_{\mathrm{IR}} \propto n_e^{3/2}$ behavior is at $n_e = g/4\pi \Lambda^2$, where $\mu_{\mathrm{IR}}$ reaches its minimum
\begin{align}\label{eq:mu_IR_min}
    \mu_{\mathrm{IR},\min} \approx \frac{e\tau_0}{m^{\star}} \frac{q_{TF}^2 \Lambda^2}{4} = \frac{ e}{8 \pi^4 \hbar} \qty(\frac{m^{\star}_z}{m^{\star}})^2 \frac{q_{TF}^2 w^6}{\Delta^2}.
\end{align}
For the high mobility structures reported in Refs.~\cite{Esposti:2023} with $w= 7$ nm, we equate the saturation mobility value $\approx 30 \times 10^{4}$ cm$^{2}$/Vs to $\mu_{\mathrm{IR},\min}$, and obtain a rough estimate of $\Delta \approx 5$ {\AA}. This value is in reasonable agreement with the roughness height measured from scanning transmission electron microscopy (STEM) images reported in Refs.~\cite{Wuetz:2023,Esposti:2023}.

Figs.~\ref{fig:mobility_infinite_noG_all} (a) and (b) are calculated using the combination of RI and IR scattering, while (c) and (d) are calculated using the combination of BI and IR scattering.
When we fit the experimental data, we restrict the RI (BI) concentration to be the same for both the 5 nm and 7 nm quantum wells.
This restriction is based on the fact that the experimental uncertainty of mobility is relatively small $\sim$10\% for different devices with the same structure design~\cite{Wuetz:2023,Esposti:2023}, indicating that the sources of disorder should be similar from device to device.
As a result, the best-fit charge impurity concentrations are $n_r = 2.5 \times 10^{12}$ cm$^{-2}$ and $N_1= 12\times10^{15}$ cm$^{-3}$.
To explain the difference in the peak mobility between the 5 nm and 7 nm quantum wells, the roughness parameters are chosen differently within the range $\Delta = 4.5-8.0$ {\AA} and $\Lambda = 16-28$ {\AA}.
The roughness parameters that we choose are of the same order of magnitude as those used in Refs.~\cite{Hwang:2004,Tracy:2009,DC_Tsui:2012,Hwang:2014_short_range}.
From Fig.~\ref{fig:mobility_infinite_noG_all} we see that the 2-impurity model combining RI and IR scattering fits the mobility data in a wider density range $n_e = 1.8-6 \times 10^{12}$ cm$^{-2}$, while the other 2-impurity model combining BI and IR fits the data in a narrower density range $n_e = 3-6 \times 10^{12}$ cm$^{-2}$.
Therefore, RI scattering is more likely the dominant scattering mechanism for transport mobility at low densities, whereas at high densities the transport mobility is limited by IR scattering.
This behavior is similar to that in Si-SiO$_2$ MOSFET 2DEG systems, where the IR scattering is much larger leading consequently to lower maximum mobilities.

In the above calculations, we ignore the local field correction $G(q) = q/(g\sqrt{q^2 + k_F^2})$ in the dielectric function Eq.~\eqref{eq:df}. 
Results that include the local field correction are presented in Figs.~\ref{fig:mobility_infinite_G_individual} and~\ref{fig:mobility_infinite_G_all}. 
Incorporating this correction leads to a reduction in the estimated impurity concentrations and the ratio of $\tau/\tau_q$.
This can be understood because the local field correction weakens the screening at a large momentum $q\gtrsim k_F$, so the disorder is more effective in scattering the electrons. 
Given that $\tau$ is more sensitive to large-momentum scattering compared to $\tau_q$, the local field correction consequently reduces $\tau$ to a greater extent than $\tau_q$ for the same disorder configuration, leading to a smaller $\tau/\tau_q$.
To analytically incorporate the local field correction $G(q)$ into the scattering rate calculations, we evaluate the $q$ integral in Eqs.~\eqref{eq:1_tau} and~\eqref{eq:1_tauq} in an approximation where the square-root singularity at $q=2k_F$ is integrated, but the form factors and $G(q)$ are taken at $q =2k_F$. 
As a result, the expressions of charged impurity scattering rates Eqs.~\eqref{eq:rate1}, \eqref{eq:rate1q}, \eqref{eq:rate2}, \eqref{eq:rate3}, \eqref{eq:rate3q}, and \eqref{eq:rate3_low_density} should be multiplied by an additional factor $F_0(2k_F w)^2[(2k_F/q_{TF})\epsilon(2k_F)]^{-2}$, where the local field correction $G(q)$ enters the dielectric function $\epsilon(q)$ defined in Eq.~\eqref{eq:df}.
At low densities $k_F \Lambda \ll 1$, the expressions of IR scattering rates Eqs.~\eqref{eq:rate_IR_low_density} and \eqref{eq:rate_qIR_low_density} should be multiplied by $[(2k_F/q_{TF})\epsilon(2k_F)]^{-2}$.
While at high densities $k_F \Lambda \gg 1$, the typical momentum transfer is $\Lambda^{-1}$ so Eqs.~\eqref{eq:rate_IR_high_density} and \eqref{eq:rate_qIR_high_density} should be multiplied by $[(2/q_{TF}\Lambda)\epsilon(2/\Lambda)]^{-2}$.

Now we address the justification of the infinite potential well approximation employed in our calculations.
This approximation is justified under two conditions. 
The first condition is that the subband gap of the infinite potential well, $E_{sg} = 3\hbar^2\pi^2/2m_z^{\star} w^2$, should satisfy the inequality $E_F \ll E_{sg} \ll V_0$, where $E_F$ is the Fermi energy of the 2DEG and $V_0$ is the conduction band offset of the quantum well.
In the case of the 7 nm quantum well reported in Ref.~\cite{Esposti:2023}, $E_{sg} \approx 25$ meV, $E_F < 4$ meV, and $V_0 \approx 180$ meV~\cite{Stern:1992,Burkard:2023}, so the first condition $E_F \ll E_{sg} \ll V_0$ is met.
The second condition is that the asymmetric electric field from the top gate should not significantly change the effective well width.
If the electric field is strong enough to tilt the bottom of the conduction band toward a triangular potential well, then the effective well width becomes smaller.
Using the Fang-Howard wavefunction~\cite{Fang_Howard:1966}, we can roughly estimate the effective well width at the highest density where the effect of the electric field $E_{\mathrm{gate}} = 4\pi e n_e/\kappa$ is the strongest.
At the highest density $n_e = 6\times 10^{11}$ cm$^{-2}$, the estimated smallest effective well width is $w_{FH} = 6/b\approx 5.5$ nm, where $b = (48\pi m_z^{\star} e^2 n_e/\kappa \hbar^2)^{1/3}$.
Given that the estimated smallest effective well width from the Fang-Howard wavefunction, $w_{FH} \approx 5.5$ nm, does not significantly differ from the original well width $w=5-7$ nm, it suggests that the bending of the conduction band bottom due to the electric field from the top gate should not substantially alter the results. 
This reasoning justifies the use of the infinite potential well approximation in our calculations.
One shortcoming of both the strict 2D and the infinite square well potential is that they cannot incorporate the short-range alloy disorder scattering arising from the SiGe alloy on the barrier outside the well.  
We address this shortcoming in the next section by using a finite well confinement model where the carrier wavefunction is allowed to tunnel into the barrier regime.

\begin{figure*}[t]
    \centering
    \includegraphics[width = 0.8\linewidth]{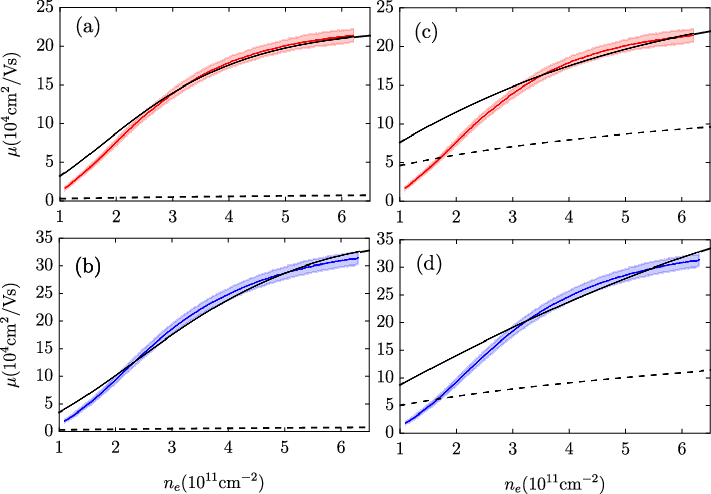}%mu_low_ne0.eps
    \caption{Mobility ($\mu$) versus electron density ($n_e$) for quantum wells with widths of 5 nm (red) and 7 nm (blue). The experimental data are depicted by red and blue curves, with corresponding error bars shown as shaded areas. The solid black curves represent the theoretical transport mobility, while the dashed black curves denote the quantum mobility, both calculated using the finite potential well approximation without the local field correction. Subfigures (a) and (b) incorporate RI, IR, and AD scattering, whereas (c) and (d) consider BI, IR, and AD scattering. Fitting parameters: (a) $n_r = 2.4 \times 10^{12}$ cm$^{-2}$, $\Delta = 4.5$ {\AA}, and $\Lambda = 50$ {\AA}; (b) $n_r = 2.4 \times 10^{12}$ cm$^{-2}$, $\Delta = 9$ {\AA}, and $\Lambda = 37$ {\AA}; (c) $N_1= 11\times10^{15}$ cm$^{-3}$, $N_2=0$, $\Delta = 5$ {\AA}, and $\Lambda = 26$ {\AA}; (d) $N_1= 11\times10^{15}$ cm$^{-3}$, $N_2=0$, $\Delta = 7$ {\AA}, and $\Lambda = 30$ {\AA}.}
    \label{fig:mobility_finite_noG_all}
\end{figure*}
\begin{figure*}[t]
    \centering
    \includegraphics[width = 0.8\linewidth]{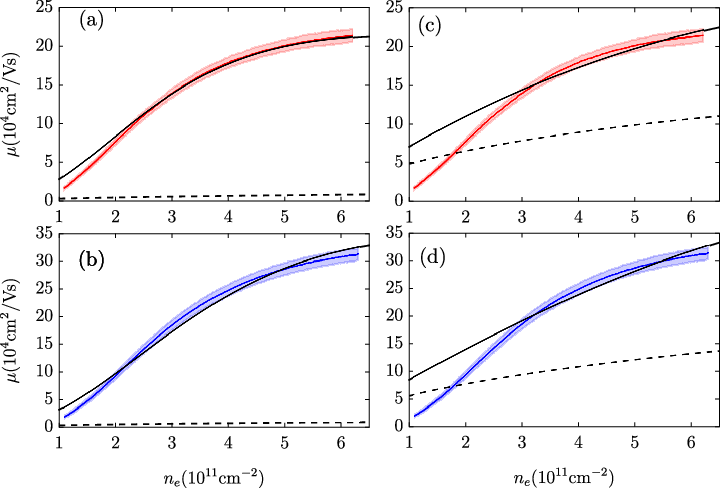}%mu_low_ne0.eps
    \caption{Mobility ($\mu$) versus electron density ($n_e$) for quantum wells with widths of 5 nm (red) and 7 nm (blue). The experimental data are depicted by red and blue curves, with corresponding error bars shown as shaded areas. The solid black curves represent the theoretical transport mobility, while the dashed black curves denote the quantum mobility, both calculated using the finite potential well approximation with the local field correction. Subfigures (a) and (b) incorporate RI, IR, and AD scattering, whereas (c) and (d) consider BI, IR, and AD scattering. Fitting parameters: (a) $n_r = 1.9 \times 10^{12}$ cm$^{-2}$, $\Delta = 4.5$ {\AA}, and $\Lambda = 22$ {\AA}; (b) $n_r = 1.9 \times 10^{12}$ cm$^{-2}$, $\Delta = 8$ {\AA}, and $\Lambda = 35$ {\AA}; (c) $N_1= 7.5\times10^{15}$ cm$^{-3}$, $N_2=0$, $\Delta = 5$ {\AA}, and $\Lambda = 20$ {\AA}; (d) $N_1= 7.5\times10^{15}$ cm$^{-3}$, $N_2=0$, $\Delta = 7.5$ {\AA}, and $\Lambda = 24$ {\AA}.}
    \label{fig:mobility_finite_G_all}
\end{figure*}

\begin{figure}[t]
    \centering
    \includegraphics[width = 0.9\linewidth]{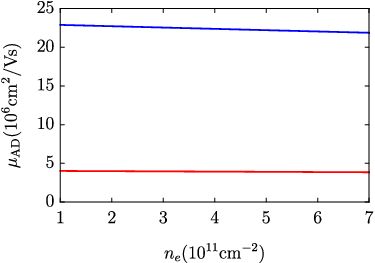}%mu_low_ne0.eps
    \caption{Mobility and quantum mobility due to alloy disorder $\mu_{\mathrm{AD}}$ versus electron density $n_e$ for a $w=5$ nm (red) and $w=7$ nm (blue) quantum well, calculated using Eq.~\eqref{eq:rate_alloy} with the finite potential well approximation.} %We ignored the dielectric screening in the calculations following Refs.~\cite{Bastard:1983,Bastard:1988}, so that mobility and the quantum mobility are equal to each other. Including the screening will make the mobility and quantum mobility values higher.}
    \label{fig:mobility_AD}
\end{figure}

\section{Finite potential well}
\label{sec:finite}
In this section, to calculate the short-range alloy disorder scattering rate, we employ the finite potential well approximation to determine the wavefunction tail extending outside the quantum well. 
Furthermore, we recalculate the RI, BI and IR scattering rates using the finite potential well wavefunction, which enters the expressions of the form factors Eqs.~(\ref{eq:df},~\ref{eq:fcq}) and the lowest subband energy in Eq.~\eqref{eq:usq2}.
The comprehensive results combining the scattering mechanisms of RI (BI), IR and AD are shown in Figs.~\ref{fig:mobility_finite_noG_all} and \ref{fig:mobility_finite_G_all}.
The finite potential well Schr{\"o}dinger equation reads
\begin{equation}\label{eq:schrodinger}
    - \frac{\hbar^2}{2 m(z)} \frac{d^2\psi(z)}{dz^2} + V(z) \psi(z) = E \psi(z)\,, 
\end{equation}
where $m(z) = m^\star_z = 0.92\,m_0$ is the longitudinal effective mass for silicon at $\abs{z} < w/2$ and $m(z) = m_B \simeq m^\star_z$ at $\abs{z} > w/2$ inside the Si$_{1-x}$Ge$_{x}$ barrier, since at $x<0.85$, Si$_{1-x}$Ge$_{x}$ alloys are considered as ``Si-like'' material with almost the same effective masses~\cite{Schaffler:1997,Schaffler:2001}.
Note that Eq.~\eqref{eq:schrodinger} is an approximation for the situation with a variable effective mass, which applies very well in our system since $m(z)$ varies little with $z$.
The effective finite potential well is described by
\begin{equation}
    V(z) = 
    \begin{cases}
        0\qc \, & \abs{z} < w/2\,,\\
        V_0 - E_F\qc \, & \abs{z} > w/2\,,
    \end{cases}
\end{equation}
where $V_0 = 180$ meV~\cite{Stern:1992,Burkard:2023} and $E_F = \hbar^2 k_F^2/2m^{\star}$ is the Fermi energy of the 2DEG.
In the following we use $V$ to represent $V_0 - E_F$ for simplicity.
%$x < 0.45$ and the energy minimum stays in the $\Gamma$ band.
%depends on the Aluminum fraction $m(x) = (0.067 + 0.083x) m_0$\MZ{This gives $m = 0.077\,m_0$}, where $m_0 = 0.511$ MeV c$^{-2}$. 
%The solution for this Schr{\"o}dinger equation can be found in Supplementary Materials.
The bound state solutions ($E<V$) for the lowest subband are
\begin{equation}\label{eq:eigenfunction}
\psi(z) =
    \begin{cases}
        C e^{-\eta \abs{z}}, \, & \abs{z}>w/2\,, \\
        D \cos(kz), \, & \abs{z}<w/2\,, \\
    \end{cases}
\end{equation}
where $k = \sqrt{2m^{\star}_z E/\hbar^2}$ and $\eta = \sqrt{2 m_B (V - E)/\hbar^2}$.
%At $z=\pm w/2$, $\psi(z)$ is continuous while $\psi'(z)$ has a finite discontinuity because of the difference in effective masses: 
The boundary condition reads
\begin{gather}
    C e^{-\eta w/2} = D \cos(kw/2)\,, \label{eq:help}\\
    \frac{C}{m_B} \eta e^{-\eta w/2} = \frac{D}{m^{\star}_z} k \sin(kw/2)\,.\label{eq:help2}
\end{gather}
The normalization of $\psi(z)$ gives
\begin{equation}
    C^2 \int_{w/2}^{\infty} e^{-2\eta z} dz + D^2 \int_0^{w/2} \cos^2(kz) dz = 1/2.
\end{equation}
Combining with Eq.~\eqref{eq:help}, one has
\begin{equation}
    C = \qty[\frac{1}{\eta} e^{-\eta w} + \frac{kw + \sin(kw)}{2k \cos^2(kw/2)} e^{-\eta w}]^{-1/2}. \label{eq:C}
\end{equation}
Dividing Eq.~\eqref{eq:help2} by Eq.~\eqref{eq:help}, one obtains
\begin{equation}\label{eq:bound_state}
    \tan(kw/2) = \frac{m^{\star}_z}{m_B} \frac{\eta}{k}\,.
\end{equation}
Introducing dimensionless quantities $\tilde E = E/E_0$, $\tilde V = V/E_0$, where $E_0 = \hbar^2 \pi^2/2 m^{\star}_z w^2$ is the ground state energy for infinite well, we arrive at~\footnote{See for example, Eq.~(5c) in Ref.~\onlinecite{gold:1989b} or Section 4.9 in Ref.~\onlinecite{davies:1997}.}
\begin{equation}\label{eq:energy}
    \tilde E + \frac{m_B}{m^{\star}_z} \tilde E \tan^2{\qty(\sqrt{\tilde E} \pi /2)} = \tilde V\,.
\end{equation}
%which justifies our infinite-potential-well approximation at lower $n_e$.
%On the other hand, at large electron concentration $n_e \in (5,15) \times 10^{11}$ cm$^{-2}$, $\tilde V$ decreased significantly from 6.2 to 1.1, and therefore finite-potential-well effect such that $0< \tilde E < 1$ has to be taken into account.
The local energy fluctuation due to interface roughness reads
\begin{equation}
    \delta E (\vb{r}) = \pdv{E}{w} \Delta(\vb{r})\,,
\end{equation}
where $\Delta(\vb{r})$ is the local variation of the well width at a position $\vb{r}$.
Taking the derivative with respect to $w$ for both sides in Eq.~\eqref{eq:energy}, we obtain the expression for $\pdv*{E}{w}$
%~\footnote{Eq.~(9) in Ref.~\onlinecite{li:2005} corresponds to $m_B =m^{\star} $ and $g(1,y) = 1$ in Eq.~\eqref{eq:dEdw}.}
\begin{equation}\label{eq:dEdw}
    \pdv{E}{w} = - \frac{2E}{w} \qty{1 + g\qty(\frac{m_B}{m^{\star}_z}, \frac{E}{V}) {\tilde V}^{-1/2} }^{-1}\,,
\end{equation}
where 
\begin{equation}
    g(x,y) =\frac{2}{\pi}\qty[x^{-1/2} + y(x^{1/2} - x^{-1/2})]^{-1}(1 - y)^{-1/2}\,.
\end{equation}
Substituting Eq.~\eqref{eq:dEdw} into Eq.~\eqref{eq:usq2}, we obtain the scattering potential for the interface roughness for a finite potential well.

We now estimate the correction introduced by the finite potential well model relative to the infinite potential well model. 
According to Eq.~\eqref{eq:energy}, we see that if $\tilde V \to \infty$ and $m_B = m^{\star}_z$, then $\tilde E = 1$. 
For finite values of $\tilde V$, we have $0< \tilde E < 1$, implying that the effective well width is larger than $w$ due to wavefunction penetration into the barriers.
For example, at electron density $n_e = 6 \times 10^{11}$ cm$^{-2}$, we have $\tilde V \approx 10$ for a 5 nm QW and $\tilde V \approx 20$ for a 7 nm QW. 
In the limit $\tilde V \gg 1$ one can solve Eq.~\eqref{eq:energy} perturbatively and obtain 
\begin{align}
    \tilde E \approx 1 - \frac{4}{\pi \tilde V^{1/2}} + \frac{12}{\pi^2 \tilde V} + \mathcal{O}\qty(\tilde V^{-\tfrac{3}{2}}),
\end{align}
so that $\tilde E \approx 0.70$ for a 5 nm QW and $\tilde E \approx 0.77$ for a 7 nm QW.
The reduced $\tilde E$ enters directly into the interface roughness scattering potential Eq.~\eqref{eq:usq2} and \eqref{eq:dEdw} and lowers the scattering rate by a factor of $\tilde E^2$.
The parentheses in Eq.~\eqref{eq:dEdw} lead to a further reduction in the scattering rate. 
As a result, $\tau_{\mathrm{IR,inf}}/\tau_{\mathrm{IR,fin}} \approx 0.34$ for a 5 nm QW and $\tau_{\mathrm{IR,inf}}/\tau_{\mathrm{IR,fin}} \approx 0.45$ for a 7nm QW.
To account for this reduction in the scattering rate due to the finite well potential, the product of $\Delta \Lambda$ must be increased to fit the mobility data, as evident from Eq.~\eqref{eq:usq2}. 
Above, we see that the finite potential well approximation significantly alters the interface roughness scattering. 
The leading order correction to $\tau_{\mathrm{IR}}$ is proportional to $\tilde V^{-1/2}$, which is substantial given that $\tilde V$ is around 10 for a 5 nm QW and 20 for a 7 nm QW.
Next, we consider how the finite potential well model alters the background impurity scattering rates. 
The model alters the form factors as per Eqs.~\eqref{eq:u_i1_mu} and \eqref{eq:df}. 
Since the form factors are proportional to the electron density $\abs{\psi(z)}^2$, the leading correction should be comparable to the leakage probability outside the quantum well
\begin{align}
    \mathrm{Prob}_{\abs{z}>w/2} = \int \limits_{\abs{z}>w/2} \abs{\psi(z)}^2 dz = C^2\eta^{-1} e^{-\eta w}.
\end{align}
In the limit of $\tilde V \gg 1$, we have
\begin{align}
    \mathrm{Prob}_{\abs{z}>w/2} \approx \frac{2}{\pi \tilde V^{3/2}}.
\end{align}
Hence, $\mathrm{Prob}_{\abs{z}>w/2} \approx 0.02$ for a 5 nm QW and $\mathrm{Prob}_{\abs{z}>w/2} \approx 0.007$ for a 7 nm QW.
This indicates that the finite potential well approximation only marginally modifies the background impurity scattering rates, affecting them by at most a few percent.

Alloy disorder within Si$_{1-x}$Ge$_x$ barriers is another short-range scattering mechanism that contributes equally to the transport and quantum scattering rate~\cite{Bastard:1983,Laikhtman:1993,Huang:2022a}
\begin{align}\label{eq:rate_alloy}
    \frac{1}{\tau_{\mathrm{AD}}} = \frac{m^{\star}}{\hbar^3} (\Delta E_c)^2 \Omega x (1-x)  \int \limits_{\abs{z}>w/2} \abs{\psi(z)}^4 dz
\end{align}
where $\Delta E_c \approx 0.8$ eV is the conduction band offset between Si and Ge~\cite{Venkataraman:1993}, $\Omega = a^3/4$ is the scatter volume with $a = 5.4$ {\AA} is the lattice constant.
In the limit of $\tilde V \gg 1$, we have $\int_{\abs{z}>w/2} \abs{\psi(z)}^4 dz \approx 2\tilde V^{-5/2}/\pi w$,
%\begin{align}
%    \int \limits_{\abs{z}>w/2} \abs{\psi(z)}^4 dz \approx \frac{2}{\pi w} \tilde V^{-5/2}.
%\end{align}
and the corresponding mobility is given by
\begin{align}\label{eq:mu_AD_estimate}
    \mu_{\mathrm{AD}} \approx \frac{w^6 e\qty(2m^{\star}_z V )^{5/2}}{2 \pi^4 \hbar^2 m^{\star2} (\Delta E_c)^2 \Omega x (1-x)} 
\end{align}
For a $w=5$ nm quantum well, $\mu_{\mathrm{AD}} \approx 1.4 \times 10^6$ cm$^2$/Vs, while for a $w=7$ nm quantum well, $\mu_{\mathrm{AD}} \approx 1 \times 10^7$ cm$^2$/Vs, which are much larger than the peak mobility reported in Refs.~\cite{Wuetz:2023,Esposti:2023}. 
The complete numerical results for alloy disorder scattering using the full finite potential well approximation are shown in Fig.~\ref{fig:mobility_AD}, and we see that the order of magnitude agrees with the rough analytical estimation of Eq.~\eqref{eq:mu_AD_estimate}.
Therefore, AD scattering is negligible in the high-mobility quantum wells reported in Refs.~\cite{Wuetz:2023,Esposti:2023}, where there is no Ge present inside the well.
(For other devices that incorporate Ge concentration within the quantum well~\cite{Eriksson:2022}, AD scattering is much stronger, resulting in an electron mobility of $2-3\times 10^4$ cm$^{2}$/Vs.)
Regarding Eq.~\eqref{eq:rate_alloy}, we follow the approach of Refs.~\cite{Bastard:1983,Bastard:1988}, omitting the dielectric function due to the short-range nature of AD around a lattice constant. 
Though several other studies have included the dielectric function in AD scattering calculations~\cite{Ando:1982a,Gold:1988,Hwang:2014_short_range}, the inclusion of dielectric screening would only increase mobility, further diminishing the role of AD. 
Therefore, our conclusion that AD scattering is not a limiting factor in the high-mobility structures reported in Refs.~\cite{Wuetz:2023,Esposti:2023} remains valid.

The results of mobility and quantum mobility calculated using the finite potential well approximation without the local field correction are shown in Fig.~\ref{fig:mobility_finite_noG_all}.
Here, Figs.~\ref{fig:mobility_finite_noG_all} (a) and (b) are calculated using a combination of RI, IR, and AD scattering. (c) and (d) are calculated using a combination of BI, IR, and AD scattering.
%The IR contributions to Fig.~\ref{fig:mobility_finite} (e) and (f) are shown separately in Fig.~\ref{fig:mobility_finite_IR}.
%The AD contributions to Figs.~\ref{fig:mobility_finite} (e) and (f) are shown separately in Fig.~\ref{fig:mobility_AD}.
The corresponding results calculated with local field correction are shown in Fig.~\ref{fig:mobility_finite_G_all}.
\begin{figure*}[t]
    \centering
    \includegraphics[width = 0.8\linewidth]{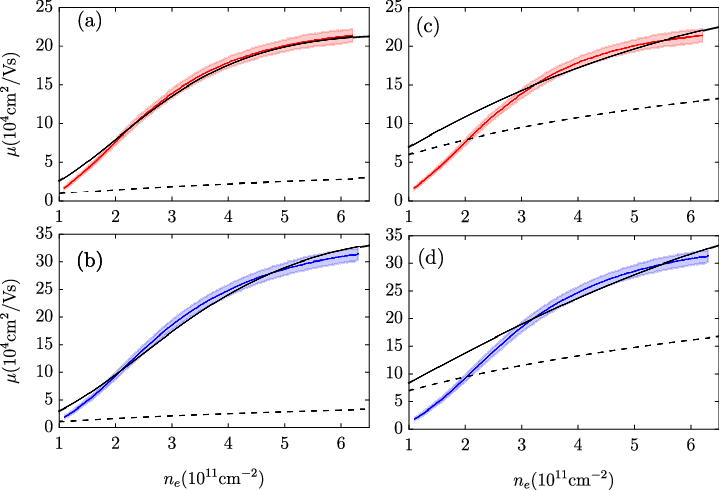}%mu_low_ne0.eps
    \caption{Mobility ($\mu$) versus electron density ($n_e$) for quantum wells with widths of 5 nm (red) and 7 nm (blue). The experimental data are depicted by red and blue curves, with corresponding error bars shown as shaded areas. The solid black curves represent the theoretical transport mobility, while the dashed black curves denote the quantum mobility, both calculated using the finite potential well approximation with gate screening and the local field correction. Subfigures (a) and (b) incorporate RI, IR, and AD scattering, whereas (c) and (d) consider BI, IR, and AD scattering. Fitting parameters: (a) $n_r = 5.2 \times 10^{12}$ cm$^{-2}$, $\Delta = 4.5$ {\AA}, and $\Lambda = 39$ {\AA}; (b) $n_r = 5.2 \times 10^{12}$ cm$^{-2}$, $\Delta = 8$ {\AA}, and $\Lambda = 36$ {\AA}; (c) $N_1= 7.5\times10^{15}$ cm$^{-3}$, $N_2=0$, $\Delta = 5$ {\AA}, and $\Lambda = 20$ {\AA}; (d) $N_1= 7.5\times10^{15}$ cm$^{-3}$, $N_2=0$, $\Delta = 7.5$ {\AA}, and $\Lambda = 24$ {\AA}.}
    \label{fig:mobility_finite_G_all_gate}
\end{figure*}
\section{Gate screening}
\label{sec:gate}
In our calculations above, the screening effect of the top gate is considered primarily through the cutoff distance $d_s$ of the background impurity distribution outside the quantum well. 
In Section~\ref{sec:delta}, we argued that $d_s = d_g/2$ should be half the distance from the gate to the center of the quantum well because the charged impurities close to the gate are effectively screened by the image charges.
In particular, we assume that the gate does not screen the remote impurities.
In this section, we refine our approach by incorporating the Coulomb potential of the image charges induced by the gate and compare the mobility results of the RI and BI scattering.
The best-fit mobility and the corresponding quantum mobility curves are depicted in Fig.~\ref{fig:mobility_finite_G_all_gate}, where the fitting concentrations of charge impurities are $n_r = 5.2 \times 10^{12}$ cm$^{-2}$ and $N_1= 7.5\times10^{15}$ cm$^{-3}$.
To explain the difference in the peak mobility between the 5 nm and 7 nm quantum wells, the roughness parameters are chosen differently within the range $\Delta = 4.5-8.0$ {\AA} and $\Lambda = 20-39$ {\AA}.
Our analysis confirms that $d_s = d_g/2$ serves as a reasonably accurate approximation for the cutoff distance of the background impurity distribution within the upper SiGe barrier.
Furthermore, we observe that the gate-screened RI scattering results align more closely with the quantum mobility data compared to the RI results that disregard gate screening. 
Previous results shown in Fig.~\ref{fig:mobility_finite_G_all} indicate that at a high density of $n_e = 6 \times 10^{11}$ cm$^{-2}$, the quantum mobility for RI without gate screening is $1.0 \times 10^4$ cm$^2$/Vs. 
This value is smaller than the experimental data point $(3.0 \pm 0.5) \times 10^4$ cm$^2$/Vs~\cite{private_comm}.
On the other hand, as illustrated in Figs.~\ref{fig:mobility_finite_G_all_gate}, when gate screening is included, the quantum mobility for RI increases to $3.0 \times 10^4$ cm$^2$/Vs at $n_e = 6 \times 10^{11}$ cm$^{-2}$, aligning well with the experimental data.
This enhancement in quantum mobility due to the inclusion of gate screening can be attributed to the suppression of small momentum forward scattering by image charges, consequently reducing the ratio $\tau/\tau_q$. 
Therefore, for a given best-fit mobility curve, the corresponding quantum mobility is expected to be higher.

In the scattering rate calculations, the gate screening is taken care by including the image charge potential into the remote impurity potential Eq.~\eqref{eq:U1_inf} at $\abs{z} > w/2$:
\begin{align}\label{eq:U1_inf_gate}
    {U}_{1}(q,z) = \frac{2\pi e^2 F_0(qw)}{\kappa q \epsilon(q)}  \qty(e^{-q\abs{z}} - e^{-q\abs{2d_g - z}}),
\end{align}
where $2d_g = 2d + 2d_o + w$.
As a result, the RI scattering rates Eqs.~\eqref{eq:rate1} and \eqref{eq:rate1q} are replaced by~\footnote{We assume the delta-layer approximation $w=0$ to get the analytical expressions.}
%[1-G(2k_F)]^2
\begin{gather}\label{eq:rate1_gate}
    \frac{1}{\tau_{\mathrm{RI}}} =  \frac{\pi\hbar n_r }{8m^{\star} k_F^3 d^3 } \qty(1 + \frac{d^3}{(2d_g - d)^3} - \frac{2d^3}{d_g^3}) \qty(\frac{2}{g})^2 , \\
    \frac{1}{\tau_{q\mathrm{RI}}} =  \frac{\pi\hbar n_r }{2 m^{\star}k_F d } \qty(1 + \frac{d}{2d_g - d} - \frac{2d}{d_g}) \qty(\frac{2}{g})^2 .\label{eq:rate1q_gate}
\end{gather}
The ratio of the scattering rates is given by
\begin{align}\label{eq:ratio_gate}
    \qty(\frac{\tau}{\tau_q})_{\mathrm{RI}} = (2k_F d)^2 I_g(d,d_g),
\end{align}
where the dimensionless function $I_g(d,d_s)$ reads
\begin{align}
    I_g(d,d_g) = \frac{1 + \frac{d}{2d_g - d} - \frac{2d}{d_g}}{1 + \frac{d^3}{(2d_g - d)^3} - \frac{2d^3}{d_g^3}}.
\end{align}
A comparison between Eqs.~\eqref{eq:ratio_RI} and \eqref{eq:ratio_gate} reveals that the inclusion of gate screening modifies the ratio by a factor of $I_g(d,d_s)$, which is independent of $k_F$.
For the high mobility structures reported in Refs.~\cite{Wuetz:2023,Esposti:2023}, $d_g \approx 44$ nm and $d \approx 30$ nm, which gives $I_g(d,d_g) \approx 0.3$.
Thus, incorporating gate screening leads to a decrease in the ratio $\qty(\tau/\tau_q)_{\mathrm{RI}}$, and consequently, the corresponding quantum mobility should increase by a factor of 3.
This explains the difference in the quantum mobility values between Figs.~\ref{fig:mobility_finite_G_all} and~\ref{fig:mobility_finite_G_all_gate} (a) (b).

On the other hand, we see that the results in Figs.~\ref{fig:mobility_finite_G_all} and~\ref{fig:mobility_finite_G_all_gate} (c) (d) are very similar to each other, suggesting that our choice of $d_s = d_g/2$ describes the gate-screened BI scattering within the upper SiGe barrier reasonably well.
We comment on the cutoff distance in the lower SiGe barrier in the context of the high-mobility structures reported in Refs.~\cite{Wuetz:2023,Esposti:2023}.
When we discuss the gate screening effects for BI in Section~\ref{sec:delta}, we assume that the cutoff distances are the same from both sides of the SiGe barriers, although in experiments the heterostructure is not symmetric. 
The bottom barrier (a few $\mu$m) is significantly thicker than the upper spacer (30 nm). 
Given that the top gate can effectively screen charge impurities only within the upper SiGe barrier, the cutoff distance for the lower SiGe barrier remains an open question.
However, as indicated by Eq.~\eqref{eq:rate3q}, at high densities, the cutoff distance $d_s$ only logarithmically changes the quantum mobility, without significantly affecting the transport mobility. 
Therefore, the choice of the cutoff distance should not change the fit to the experimental mobility data. 
By considering the cutoff distance as an additional fitting parameter, one can always adjust the model to match a specific quantum mobility data point.
A more comprehensive future experimental investigation of high-mobility Si quantum wells would be instrumental in enhancing our understanding of the BI distribution within the SiGe barrieres.
%Therefore, although the results of gate-screened RI scattering itself agree with the experimental data point at high density, one cannot rule out the possibility

In summary, the results of our model, which incorporates gate screening, suggest that RI scattering is likely the predominant scattering mechanism for transport mobility at low densities, whereas at high densities the transport mobility is restricted by IR scattering.
The quantum mobility predicted for gate-screened RI scattering aligns closely with experimental data within the margin of error at the highest density examined. 
This observation suggests that RI scattering could be the limiting factor for quantum mobility across all densities.
On the other hand, since the cutoff distance of the BI distribution inside the bottom SiGe barrier remains an open question, we cannot rule out the possibility that some distant background impurities also contribute to the quantum mobility.
Future experiments focusing on the quantum mobility over a broader range of densities could yield more quantitative information on the concentration and the cutoff distance of background impurities inside the bottom SiGe barrier.
%Such studies would be invaluable in advancing our understanding of scattering mechanisms and their impact on mobility in high-mobility Si quantum wells.
%On the other hand, we should again emphasize that the experimental accuracy of the quantum mobility data in general is much less reliable than that for the transport mobility~\cite{Stern:1985}.

%Using our theoretical models to interpret the quantum mobility data can only suggest 

\begin{table}
\caption{Critical density $n_c$ and $n_{cq}$ in units of $10^{11}$ cm$^{-2}$ for different quantum degeneracy $g = 4$, 2, and 1. The results are obtained from $k_F l = 1$ and $k_F l_q = 1$ using the same impurity parameters from Fig.~\ref{fig:mobility_finite_G_all_gate} calculated in a finite potential well approximation with gate screening and local field correction. RI (BI) indicates the dominant scattering mechanism at low densities. 5 nm (7 nm) means the calculation for a 5 nm (7 nm) QW.}
\begin{ruledtabular}
\begin{tabular}{c c c c c} 
 & RI (5 nm) & RI (7 nm) & BI (5 nm) & BI (7 nm)\\
%heading
%\hline \\ [-2.5ex]
$n_c$ (g=4) & 0.52 & 0.49 & 0.15 & 0.14  \\
$n_{cq}$ (g=4) & 0.65 & 0.62 & 0.12 & 0.11  \\
\hline \\ [-2.5ex]
$n_c$ (g=2) & 0.50 & 0.48 & 0.28 & 0.26  \\
$n_{cq}$ (g=2) & 0.83 & 0.79 & 0.28 & 0.25  \\
\hline \\ [-2.5ex]
$n_c$ (g=1) & 0.49 & 0.47 & 0.42 & 0.38  \\
$n_{cq}$ (g=1) & 1.1 & 1.0 & 0.48 & 0.43  \\
\end{tabular}
\end{ruledtabular}
%\vspace{-0.2 in}
\label{table:nc_air}
\end{table}

\begin{table}
\caption{Critical density $n_c$ in units of $10^{11}$ cm$^{-2}$ obtained by the power law fit $\sigma \propto (n_e - n_c)^p$ to the conductivity data $\sigma = e n_e \mu$ at low densities $n_e =1-2 \times 10^{11}$ cm$^{-2}$ reported in Refs.~\cite{Wuetz:2023,Esposti:2023}. The parenthesis show the 95\% confidence interval of the fitting parameters. The first column fixed $p=1.31$ assuming the 2D percolation. The second and third columns are results treating $p$ as a free parameter.}
\begin{ruledtabular}
\begin{tabular}{c c c c c} 
 & $n_c$ ($p=1.31$) & $n_c$ & $p$ \\
%heading
%\hline \\ [-2ex] 
5 nm & (0.837, 0.855) & (0.436, 0.514) & (2.20, 2.41) \\ 
7 nm & (0.869, 0.889) & (0.447, 0.515) & (2.34, 2.52) \\ 
\end{tabular}
\end{ruledtabular}
%\vspace{-0.2 in}
\label{table:nc_data_fit}
\end{table}

\section{Metal insulator transition}
\label{sec:mit}
In this section, we discuss the mobility data in the low-density regime, and estimate the critical density of the metal-insulator transition (MIT) using the Anderson-Ioffe-Regel (AIR) condition~\cite{anderson1958,ioffe1960}.
All our results in the previous sections are calculated using the Boltzmann-Born transport theory, which is valid at high densities such that $k_F l \gg 1$, where $l=v_F \tau$ is the transport mean free path.
One might expect that the Boltzmann-Born theory fails at a critical density $n_e = n_{c}$ where $k_F l = 1$.
This is the well-known AIR criterion of MIT.
However, deviations in the metallic regime of our results, as shown in Fig.~\ref{fig:mobility_finite_G_all_gate}, suggest that the theoretical mobility curves, fitting well at high densities, start overestimating mobility at low densities compared to the experimental data.
%However, from our results in Fig.~\ref{fig:mobility_finite_G_all_gate}, we see that the best-fit theoretical mobility curves start to deviate and are systematically higher than the experimental data at low densities.
It is crucial to emphasize that these low-density deviations are not due to any neglected scattering mechanism in the theory. 
%We should again emphasize that this deviation at low densities does not arise from any new scattering mechanism ignored in the theory.
%The dominant scattering mechanism at low densities is remote charged impurity.
The dominant scattering mechanism at low densities is RI (or BI, depending on the choice of the impurity model). 
For RI-dominant cases, as depicted in Figs.~\ref{fig:mobility_finite_G_all_gate} (a) and (b), the deviation is observed at $n_e \lesssim 1.8 \times 10^{11}$ cm$^{-2}$, where $k_F l \lesssim 30$ or $k_F l_q \lesssim 6$ with $l_q = v_F \tau_q$~\cite{Ahn_MIT_Si:2022}. 
On the other hand, for BI dominant scenarios, shown in Figs.~\ref{fig:mobility_finite_G_all_gate} (c) and (d), the deviation occurs at a higher density $n_e \lesssim 2.8 \times 10^{11}$ cm$^{-2}$, around $k_F l \lesssim 100$ or $k_F l_q \lesssim 60$.
%For the RI dominant case shown in Figs.~\ref{fig:mobility_finite_G_all_gate} (a) and (b), this deviation occurs at $n_e \lesssim 1.8 \times 10^{11}$ cm$^{-2}$ around $k_F l \lesssim 30$ or $k_F l_q \lesssim 6$ with $l_q = v_F \tau_q$~\cite{Ahn_MIT_Si:2022}.
%While for the BI dominant case shown in Figs.~\ref{fig:mobility_finite_G_all_gate} (c) and (d), this deviation occurs at a higher density $n_e \lesssim 2.8 \times 10^{11}$ cm$^{-2}$ around $k_F l \lesssim 100$ or $k_F l_q \lesssim 60$. %where the Boltzmann-Born theory should still be applicable since $k_F l \gg 1$. 
%The reason for the earlier breakdown of the Boltzmann-Born theory is as follows.
%As a first thought, it is expected that the semiclassical Boltzmann-Born transport theory should get worse when $k_F l$ becomes small, and the low density result should be modified by more higher order scattering processes.
%The reasons for this breakdown of the Boltzmann-Born theory are as follows.
%Due to long-range Coulomb disorder, the Boltzmann-Born theory fails at sufficiently low densities, where the 2DEG is broken into puddles separated by disorder potential barrier with a highly inhomogeneous landscape.
The failure of the Boltzmann-Born theory at low densities is due to the following. 
Long-range Coulomb disorder leads to the fragmentation of the 2DEG into electron puddles, separated by disorder potential barriers, creating an inhomogeneous 2DEG landscape. 
%Although $k_F l \gg 1$ is still locally satisfied inside each electron puddle, the linear screening theory assuming a uniform 2EDG is no longer applicable.
Although $k_F l \gg 1$ is locally satisfied within each puddle, the uniform 2DEG assumption, fundamental to linear screening theory, is no longer valid. 
Instead, transport properties in the low-density region $n_e \geq n_c$ near MIT should be described by percolation through charged puddles, following $\sigma \propto (n_e - n_c)^p$ with $p=1.31$~\cite{Thouless:1971,Kirkpatrick:1973,Shklovskii:1975,DasSarma:2005579,Tracy:2009,Manfra:2007,Qiuzi:2013,Tracy:2014}, and cannot be captured by the Boltzmann-Born theory even including all high-order scattering processes.
The failure of the Boltzmann-Born theory at low densities also indicates that the critical density estimated from the AIR criterion should be smaller than the actual critical density (percolation threshold) observed in experiments and should be used as a lower bound estimation of the percolation threshold.

%Using $N_1 \approx 5\times 10^{15}$ cm$^{-3}$ and $a_B = 3.3$ nm for Si, we get $n_{c {\rm ES}} \approx 1.2\times10^{11}$ cm$^{-2}$.
%Although this result is reasonably in agreement with the percolation density $7 - 8 \times 10^{10}$ cm$^{-2}$ reported in Refs.~\cite{Wuetz:2023,Esposti:2023}.

Nevertheless, we apply the AIR criterion, characterized by $k_{F} l = 1$ and $k_{F} l_q = 1$, to estimate the critical densities $n_c$ and $n_{cq}$~\cite{anderson1958,ioffe1960}. 
Utilizing the delta-layer 2DEG approximation, as discussed in Sections~\ref{sec:delta} and \ref{sec:gate}, we can analytically estimate the critical density of the metal-insulator transition (MIT). 
By substituting the gate-screened remote impurity scattering rates from Eqs.~\eqref{eq:rate1_gate} and \eqref{eq:rate1q_gate} into the AIR criterion, we derive:
%\qty(1-\frac{2}{g\sqrt{5}})^{4/5}} \qty(1-\frac{2}{g\sqrt{5}})^{4/3}
\begin{gather}\label{eq:nc_RI}
    n_{c}^{\mathrm{(RI)}} = \frac{g^{1/5}\qty[1 + \frac{d^3}{(2d_g - d)^3} - \frac{2d^3}{d_g^3}]^{2/5} }{2^{12/5} \pi^{3/5} }  \qty(\frac{n_r}{d^3})^{2/5},\\
    \label{eq:ncq_RI}
    n_{cq}^{\mathrm{(RI)}} = \frac{g^{-1/3} \qty[1 + \frac{d}{2d_g - d} - \frac{2d}{d_g}]^{2/3}}{2^{4/3} \pi^{1/3} } \qty(\frac{n_r}{d})^{2/3},
\end{gather}
Eqs.~\eqref{eq:nc_RI} and \eqref{eq:ncq_RI} are applicable if $n_{c}^{\mathrm{(RI)}}$ and $n_{cq}^{\mathrm{(RI)}}$ exceed $g/16\pi d^2$. 
This condition is met in high-mobility quantum wells~\cite{Wuetz:2023,Esposti:2023} where $g/16\pi d^2 \approx 10^{10}$ cm$^{-2}$.
For BI scattering, substituting the rates from Eqs.~\eqref{eq:rate3} and \eqref{eq:rate3q} into the AIR criterion yields:
\begin{gather}\label{eq:nc_BI_high}
    n_{c}^{\mathrm{(BI)}} = g^{1/5} 2^{-8/5} \pi^{-3/5} \qty(\frac{N_1}{w^2})^{2/5},\\
    n_{cq}^{\mathrm{(BI)}} = g^{-1/3} 2^{-2/3} \pi^{-1/3} \qty[\ln(2d_s/w)]^{2/3} N_1^{2/3},\label{eq:ncq_BI_high}
\end{gather}
which are valid if $n_{c}^{\mathrm{(BI)}}$ and $n_{cq}^{\mathrm{(BI)}}$ are greater than $g/16\pi w^2$.
Otherwise, we use Eq.~\eqref{eq:rate3_low_density} and obtain
\begin{align}\label{eq:nc_BI_low}
    n_{c}^{\mathrm{(BI)}} = n_{cq}^{\mathrm{(BI)}} = \frac{\pi}{g} N_1(2d_s - w),
\end{align}
which is applicable if $n_{c}^{\mathrm{(BI)}} < g/16\pi w^2$.
Given that the crossover density $g/16\pi w^2 \approx 2.4 \times 10^{11}$ cm$^{-2}$ is significantly higher than the MIT critical density in high-mobility quantum wells~\cite{Wuetz:2023,Esposti:2023}, we should use Eq.~\eqref{eq:nc_BI_low} for estimating $n_c$ in BI-dominated scenarios.
%Since the crossover density $g/16\pi w^2 \approx 2.4 \times 10^{11}$ cm$^{-2}$ is apparently much higher than the MIT critical density in high-mobility quantum wells reported in Refs.~\cite{Wuetz:2023,Esposti:2023}, we should use Eq.~\eqref{eq:nc_BI_low} to estimate $n_c$ if the dominant scattering source is BI.
Using $d=30$ nm, $d_s = d_g/2= 22$ nm, and the impurity concentration $n_r \approx 5.2 \times 10^{12}$ cm$^{-2}$ and $N_1 \approx 7.5\times 10^{15}$ cm$^{-3}$ from Fig.~\ref{fig:mobility_finite_G_all_gate}, we obtain the critical densities $n_{c}^{\mathrm{(RI)}} \approx 5 \times 10^{10}$ cm$^{-2}$, $n_{cq}^{\mathrm{(RI)}} \approx 7 \times 10^{10}$ cm$^{-2}$, $n_{c}^{\mathrm{(BI)}} = n_{cq}^{\mathrm{(BI)}} \approx 2 \times 10^{10}$ cm$^{-2}$.
%Our analytical estimates of the critical densities are in reasonable agreement with the percolation densities $n_c = 7-8\times 10^{10}$ cm$^{-2}$ reported in Refs.~\cite{Wuetz:2023,Esposti:2023}.

Next, we estimate the critical density of MIT by solving the AIR conditions $k_{F} l = (2/g)(h n_e \mu/e) = 1$ and $k_{F} l_q = (2/g)(h n_e \mu_q/e) = 1$ with $\mu$ and $\mu_q$ numerically calculated from the best fits shown in Fig.~\ref{fig:mobility_finite_G_all_gate}.
%The complete numerical results of $n_c$ and $n_{cq}$ are summarized in the first two rows of Table~\ref{table:nc_air}, whose values are similar to the analytical results obtained in Eqs.~\eqref{eq:nc_RI}, ~\eqref{eq:ncq_RI}, and \eqref{eq:nc_BI_low}.
The complete numerical results for $n_c$ and $n_{cq}$ are summarized in the first two rows of Table~\ref{table:nc_air}. 
These values closely align with the analytical results from Eqs.~\eqref{eq:nc_RI}, \eqref{eq:ncq_RI}, and \eqref{eq:nc_BI_low}.
Furthermore, Table~\ref{table:nc_air} includes results for various quantum degeneracies $g = 4$, 2, and 1, while keeping the same disorder parameters as used in Fig.~\ref{fig:mobility_finite_G_all_gate}. 
%Furthermore, in Table~\ref{table:nc_air} we show the results of $n_c$ and $n_{cq}$ for different quantum degeneracy $g = 4$, 2, and 1, keeping the same impurity parameters used in Fig.~\ref{fig:mobility_finite_G_all_gate}.
%We see that the changes of $n_c$ and $n_{cq}$ as a function of $g$ are semi-quantitatively described by Eqs.~\eqref{eq:nc_RI}, \eqref{eq:ncq_RI}, and \eqref{eq:nc_BI_low}.
The changes in $n_c$ and $n_{cq}$ as functions of $g$ are semi-quantitatively consistent with Eqs.~\eqref{eq:nc_RI}, \eqref{eq:ncq_RI}, and \eqref{eq:nc_BI_low}. 
%For the best fits shown in Fig.~\ref{fig:mobility_finite_G_all} without taking into account gate screening, the corresponding critical densities are $n_{c}^{(\mathrm{RI})} \approx 4 \times 10^{10}$ cm$^{-2}$, $n_{cq}^{(\mathrm{RI})} \approx 8 \times 10^{10}$ cm$^{-2}$, and $n_{c}^{(\mathrm{BI})} \approx n_{cq}^{(\mathrm{BI})} \approx 1 \times 10^{10}$ cm$^{-2}$. 
%By including gate screening effects, the results of $n_c$ change by a few percent, but $n_{cq}$ is reduced by roughly a factor of 2.
%This happens because small-angle scattering is suppressed by gate screening, which increases $\tau_q$ and decreases $n_{cq}$.
We should emphasize that our estimate of $n_c$ and $n_{cq}$ using the AIR criterion shown in Table~\ref{table:nc_air} should serve as a lower bound for the experimental critical density of MIT.

Considering that the nature of 2D MIT in high-mobility Si quantum wells likely resembles a density-inhomogeneity-driven percolation transition similar to Si MOSFETs~\cite{Tracy:2009}, we performed a percolation fit of the experimental data to estimate the experimental critical density. 
Refs.~\cite{Wuetz:2023,Esposti:2023} reported the percolation density $n_c = 7\times 10^{10}$ cm$^{-2}$ ($n_c = 8\times 10^{10}$ cm$^{-2}$) for their high-mobility 7 nm (5 nm) Si quantum wells.
The percolation density can also be estimated approximately by considering inhomogeneous density fluctuations in the system~\cite{Hwang:2014a,Efros:19881019,Pikus:1989,Tracy:2009,Qiuzi:2013,Ahn_MIT:2023}, leading to $n_c \approx 0.1 \sqrt{n_r}/d \approx 7.5 \times 10^{10}$ cm$^{-2}$, where we used $d=30$ nm and $n_r \approx 5 \times 10^{12}$ cm$^{-2}$.
This $n_c$ is higher than the one obtained from the AIR criterion.
Furthermore, we individually performed the percolation fit $\sigma \propto (n_e - n_c)^p$ using the conductivity data $\sigma = e n_e \mu$ at low densities $n_e =1-2 \times 10^{11}$ cm$^{-2}$ reported in Refs.~\cite{Wuetz:2023,Esposti:2023}.
The results of the best-fit $n_c$ and $p$ are presented in Table~\ref{table:nc_data_fit}.
The percolation fit with fixed $p=1.31$ yields $n_c \approx 0.85 \times 10^{11}$ cm$^{-2}$.
Allowing $p$ as a fitting parameter results in $p\approx 2.4$ and $n_c \approx 0.5 \times 10^{11}$ cm$^{-2}$. 
Our percolation fit results are in reasonable agreement with those reported in Refs.~\cite{Wuetz:2023,Esposti:2023}.
Comparing Tables~\ref{table:nc_air} and \ref{table:nc_data_fit}, we observe that the RI results provide a more stringent lower bound of the critical density, as $n_{c}^{\mathrm{(RI)}}$ and $n_{cq}^{\mathrm{(RI)}}$ are closer to the percolation densities reported in Refs.~\cite{Wuetz:2023,Esposti:2023}.
This suggests that RI scattering, rather than BI, is likely the limiting mechanism at low densities in the studied high-mobility Si quantum wells.
Given that $n_{cq}^{\mathrm{(RI)}} > n_c^{\mathrm{(RI)}}$, we find that the AIR criterion, which includes the single-particle scattering time $\tau_q$~\cite{Ahn_MIT_Si:2022}, provides a more accurate estimate of the percolation density.
In the future, a more systematic experimental study of the temperature and density dependence of the conductivity is needed to gain a more comprehensive understanding of the 2D MIT in high-quality Si quantum wells. %in high mobility structures reported in Refs.~\cite{Wuetz:2023,Esposti:2023}.

\begin{figure*}[t]
    \centering
    \includegraphics[width = 0.8\linewidth]{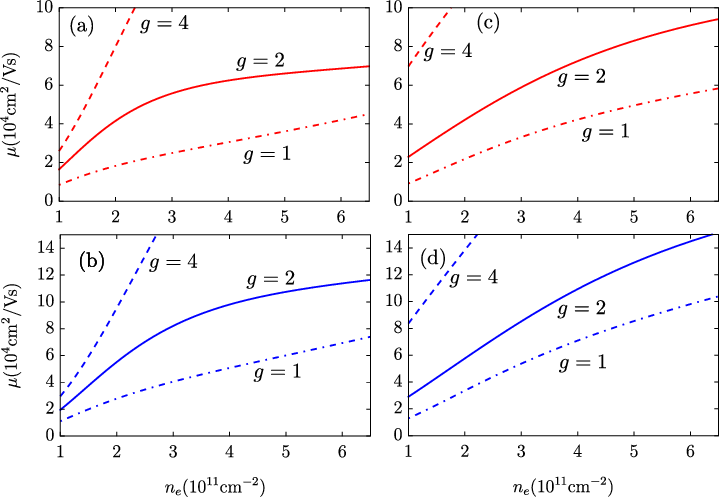}%mu_low_ne0.eps
    \caption{Mobility ($\mu$) versus electron density ($n_e$) for quantum wells with widths of 5 nm (red) and 7 nm (blue). The dashed, solid, and dot-dashed curves represent the theoretical mobility assuming the total quantum degeneracy $g=4$, 2, and 1 respectively. The calculations are done using the finite potential well approximation with gate screening and local field correction. Subfigures (a) and (b) incorporate RI, IR, and AD scattering, whereas (c) and (d) consider BI, IR, and AD scattering. The corresponding disorder parameters are the same as in Fig.~\ref{fig:mobility_finite_G_all_gate}.}
    \label{fig:mobility_finite_G_all_gate_valley_comparison}
\end{figure*}
\section{Effects of Spin/Valley degeneracy}
\label{sec:valley}
In earlier sections, our numerical calculations considered a total quantum degeneracy $g = g_s g_v = 4$.
This assumption is valid under two conditions: first, the Fermi energy must substantially surpass the valley splitting, and second, the external magnetic field should be weak enough to avoid inducing spin polarization. 
In this scenario, both the valley and spin degrees of freedom maintain a twofold degeneracy, indicated by $g_v = g_s = 2$.
This section theoretically addresses how the total quantum degeneracy $g$ affects the electron transport properties in the high-mobility Si/SiGe structures for a given disorder configuration.
For example, a decrease in degeneracy from $g=4$ to $g=2$ can be achieved by applying an in-plane magnetic field $B$ parallel to the 2DEG, which is assumed to only affect the spin degeneracy of the 2DEG without orbital effect if the thickness of the 2DEG is much smaller than the cyclotron radius $w \ll c \hbar k_F/e B$~\cite{Hwang_parallel_B_field:2000,Hwang_parallel_B_field:2005,Hwang_parallel_B_field:2005b,Hwang:2013a_valley}.
Such a reduced degeneracy is also possible without any spin polarization if the valley splitting happens to be larger than the Fermi energy so that $g_v=1$ in the system.

In the strong screening limit $q_{TF}/2k_F \gg 1$ (for the density range of interest to us $n_e \ll g^3/16 \pi a_B^2$), mobility increases monotonically with increasing $g$ for a given impurity configuration. 
This monotonic behavior of mobility versus $g$ has been shown in a comprehensive analysis of valley-dependent 2D transport in (100), (110), and (111) Si inversion layers with the same bare Coulomb disorder~\cite{Hwang:2013a_valley,Hwang:2007}. 
A similar phenomenon was observed in an ambipolar gate-controlled Si(111)-vacuum field effect transistor, where by tuning the external gate voltage the 2D electron and hole transport can be studied within the same device against the same background Coulomb disorder~\cite{Binhui:2015}. 
For the ambipolar transistor discussed in Ref.~\cite{Binhui:2015}, the effective valley degeneracy is 6 for a 2D electron system,, whereas it is 1 for a 2D hole system, resulting in the peak electron mobility being approximately 20 times greater than the peak hole mobility at low temperatures. 
This can be qualitatively understood from the expression of the Thomas-Fermi screening wave vector $q_{TF} = g/a_B$, implying that higher $g$ leads to improved screening and consequently, higher mobility.
However, the relationship between mobility and $g$ is generally nontrivial, as $g$ influences both $q_{TF} \propto g$ and $k_F \propto g^{-1/2}$.
For remote impurity scattering, it is evident from Eqs.~\eqref{eq:rate1}, \eqref{eq:rate1q}, \eqref{eq:rate1_gate}, and \eqref{eq:rate1q_gate} that $\mu_{\mathrm{RI}} \propto g^{1/2}$ and $\mu_{q\mathrm{RI}} \propto g^{3/2}$. 
On the other hand, for short-range impurity scattering, $\mu_{\mathrm{S}} = \mu_{q\mathrm{S}} \propto g^{2}$, as indicated by Eq.~\eqref{eq:rate2}. 
As discussed in Section~\ref{sec:delta}, the $g$-dependence of background impurity scattering crosses over from short-range to remote impurity scattering behavior with increasing electron density, as detailed in Eqs.~(\ref{eq:rate3}-\ref{eq:rate3_low_density}). 
The crossover density is at $n_e = g/16 \pi w^2$.
%As we explained in Section~\ref{sec:delta}, the $g$-dependence of background impurity scattering should cross over from the behavior of short-range impurity scattering to the behavior of remote impurity scattering by increasing the electron density [cf. Eqs.~(\ref{eq:rate3}-\ref{eq:rate3_low_density})].
%The crossover density is at $n_e = g/16 \pi w^2$.
For interface roughness scattering, the mobilities $\mu_{\mathrm{IR}}$ and $\mu_{q\mathrm{IR}}$ are proportional to $g^3$ at low densities $n_e < g/4\pi \Lambda^2$, transitioning to a long-range behavior where $\mu_{\mathrm{IR}} \propto g^{1/2}$ and $\mu_{q\mathrm{IR}} \propto g^{3/2}$ at higher densities, as shown in Eqs.~(\ref{eq:rate_IR_low_density}-\ref{eq:rate_qIR_high_density}). 
However, IR scattering at low densities $n_e < g/4\pi \Lambda^2$ is typically overshadowed by the dominant scattering mechanism at these densities, which is usually remote Coulomb disorder. 
At $n_e = g/4\pi \Lambda^2$, the minimum mobility due to IR scattering $\mu_{\mathrm{IR},\min} \propto g^2$, as per Eq.~\eqref{eq:mu_IR_min}.
Since the unscreened alloy disorder scattering rate, as given by Eq.~\eqref{eq:rate_alloy}, depends on $k_F$ only through $V=V_0 - E_F$, the mobility $\mu_{\mathrm{AD}}$ is not influenced by $g$ provided that $V_0 \gg E_F$ is fulfilled.
The results of the above analysis suggest that if a parallel magnetic field is used to lift the spin degeneracy without any orbital effect, the metallic resistance can at most be increased by a factor of 4~\cite{Okamoto:1999,Dolgopolov:2000,Herbut:2001,Hwang_screening:2005,Hwang_parallel_B_field:2005,Hwang_parallel_B_field:2005b}.

Fig.~\ref{fig:mobility_finite_G_all_gate_valley_comparison} presents the results of mobility $\mu(n_e)$ for different quantum degeneracies $g = 4$, 2, and 1. 
For these calculations, the finite potential well approximation with gate screening and local field correction is utilized, employing the same disorder parameters as in Fig.~\ref{fig:mobility_finite_G_all_gate}. 
As expected, we see that the mobility decreases monotonically with the reduction in $g$. 
The scattering mechanism that limits the transport mobility is IR at high densities, while the limiting factor of mobility at low densities is RI (or BI, depending on the choice of the impurity model).
The dashed curves in Fig.~\ref{fig:mobility_finite_G_all_gate_valley_comparison} replicate the $g=4$ mobility results from Fig.~\ref{fig:mobility_finite_G_all_gate}.
The solid curves represent our prediction for the mobility in the presence of a parallel magnetic field without orbital effect.
In this scenario, the spin is fully polarized ($g_s = 1$), resulting in a total quantum degeneracy of $g = g_v = 2$. 
The observed ratio of $\mu_{g=4}/\mu_{g=2}$ falls within the range from $\sqrt{2}$ to $4$, which aligns with the predicted $g^{1/2}$ to $g^{2}$ scaling of $\mu$ across various scattering mechanisms. 
Our choice of the valley degeneracy $g_v = 2$ is appropriate for the high mobility structures reported in Refs.~\cite{Wuetz:2023,Esposti:2023}, where the Fermi energy in the density range $n_e = 1-6 \times 10^{11}$ cm$^{-2}$ is $E_F \in (0.6,3.8) \times (4/g)$ meV. 
This Fermi energy is significantly greater than the valley splitting of $0.2$ meV.
Should the valleys split further such that only one valley is occupied, combined with a strong parallel magnetic field, one could achieve a quantum degeneracy of $g=1$. 
The dot-dashed curves in Fig.~\ref{fig:mobility_finite_G_all_gate_valley_comparison} are our predictions in this scenario, indicating a further reduction in mobility.

The quantum degeneracy $g$ also has an effect on the critical density $n_c$ estimated from the AIR criterion. 
The specific relationship between $n_c$ and $g$ depends on the dominant scattering mechanism responsible for MIT and the dimensionless parameters such as $q_{TF}/2k_F$, $k_F d$ for RI, $k_F w$ for BI, and $k_F \Lambda$ for IR.
These parameters determine the effective screened potential that scatters electrons, thereby altering the MIT critical density.
In the strong screening limit $q_{TF}/2k_F \gg 1$ or $n_e \ll g^3/16 \pi a_B^2$, which is of interest to us for the high-mobility structures reported in Refs.~\cite{Wuetz:2023,Esposti:2023}, the critical densities estimated from the AIR criterion for the RI and BI scatterings are summarized in Eqs.~(\ref{eq:nc_RI}-\ref{eq:nc_BI_low}).
We see that for long-range Coulomb impurity scattering, $n_c \propto g^{1/5}$ while $n_{cq} \propto g^{-1/3}$.
For MIT induced by short-range impurity scattering, the critical density is given by 
\begin{align}\label{eq:nc_short}
    n_{c}^{\mathrm{(S)}} = n_{cq}^{\mathrm{(S)}} = \frac{\pi}{g} n_2,
\end{align}
which is the same as Eq.~\eqref{eq:nc_BI_low} with $N_1(2d_s - w)$ replaced by the effective 2D short-range impurity concentration $n_2$ nearby the 2DEG.
For example, Eq.~\eqref{eq:nc_short} describes the case in Si MOSFETs where $n_c$ increases as $g$ (and $\mu$) decreases~\cite{Sarachik_2002,Hwang:2014a,Ahn_MIT_Si:2022}.
Table~\ref{table:nc_air} shows the value of $n_c$ and $n_{cq}$ for the same disorder parameters taken from Fig.~\ref{fig:mobility_finite_G_all_gate} but with different $g$.
We should emphasize that the disorder parameters in Fig.~\ref{fig:mobility_finite_G_all_gate} are obtained by fitting the experimental data with the theoretical mobility curves using $g=4$, which is justified since $E_F$ is much larger than the valley splitting in the high-mobility Si quantum wells.
For completeness, Appendix~\ref{sec:appendix} presents the best-fit results of the mobility curves and the corresponding critical densities assuming $g=2$. 
In this scenario, the fitting disorder parameters are lower than those in Fig.~\ref{fig:mobility_finite_G_all_gate} due to weaker screening.

\section{Conclusion}\label{sec:conclusion}
In summary, as the main goal of our theoretical work in the context of the importance of the Si/Ge quantum computing platform, we have systematically examined the roles of different scattering sources on the transport and quantum mobilities in the Si/Si$_{0.69}$Ge$_{0.31}$ quantum wells with a co-design of high electron mobility and large valley splitting reported in Refs.~\cite{Wuetz:2023,Esposti:2023}.
We identify remote charged impurities and interface roughness as primary limitations for transport mobility at low and high electron densities, respectively. 
Quantum mobility, on the other hand, influenced by both remote and distant background impurities, requires further experimental investigation to understand the charged impurity distribution in SiGe barriers.
Our analysis finds alloy disorder scattering to be negligible in high-mobility structures. 
Other scattering sources, such as inhomogeneous strain distributions and threading dislocations~\cite{crystal_quality}
that originated from the bottom metamorphic substrate of several $\mu$m, are also quantitatively irrelevant for the reported mobility using similar estimations from Ref.~\cite{Monroe:1993,Ismail:1994}.
We estimate the metal-insulator transition critical density using the Anderson-Ioffe-Regel criterion and qualitatively explain the breakdown of the Boltzmann-Born theory at low densities due to the long-range Coulomb disorder. 
We also estimate the critical density by considering inhomogeneous density fluctuations induced by long-range Coulomb disorder in the system and find a larger critical density compared to the one obtained from the Anderson-Ioffe-Regel criterion.
To gain a more comprehensive understanding of the metal-insulator transition, more extensive future experimental studies of the temperature and density dependence of the conductivity should be conducted.
We also predict the electron mobility with reduced quantum degeneracy, where the spin is fully polarized by an external parallel magnetic field, for future experimental validation.

\begin{acknowledgments}
The authors thank G. Scappucci and D.D. Esposti for discussions and clarification of the experimental details.
This work is supported by the Laboratory for Physical Sciences.
\end{acknowledgments}

\appendix
\section{Best-fit results assuming \texorpdfstring{$g=2$}{no_valley_degeneracy}}
\label{sec:appendix}
\begin{figure*}[t]
    \centering
    \includegraphics[width = 0.8\linewidth]{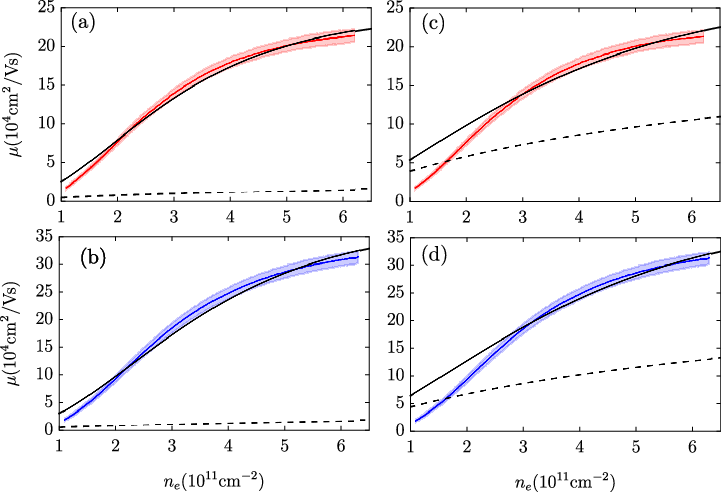}%mu_low_ne0.eps
    \caption{Mobility ($\mu$) versus electron density ($n_e$) for quantum wells with widths of 5 nm (red) and 7 nm (blue). The experimental data are depicted by red and blue curves, with corresponding error bars shown as shaded areas. The solid black curves represent the theoretical transport mobility, while the dashed black curves denote the quantum mobility, both calculated using the finite potential well approximation with gate screening and the local field correction, assuming the total quantum degeneracy $g = 2$. Subfigures (a) and (b) incorporate RI, IR, and AD scattering, whereas (c) and (d) consider BI, IR, and AD scattering. Fitting parameters: (a) $n_r = 3.6 \times 10^{12}$ cm$^{-2}$, $\Delta = 3.5$ {\AA}, and $\Lambda = 20$ {\AA}; (b) $n_r = 3.6 \times 10^{12}$ cm$^{-2}$, $\Delta = 8$ {\AA}, and $\Lambda = 15$ {\AA}; (c) $N_1= 3.3\times10^{15}$ cm$^{-3}$, $N_2=0$, $\Delta = 4$ {\AA}, and $\Lambda = 15$ {\AA}; (d) $N_1= 3.3\times10^{15}$ cm$^{-3}$, $N_2=0$, $\Delta = 6.5$ {\AA}, and $\Lambda = 18$ {\AA}.}
    \label{fig:mobility_finite_G_all_gate_gv1}
\end{figure*}

In this appendix, we present the best-fit results of mobility curves and the corresponding MIT critical densities assuming the total quantum degeneracy $g=2$.
Fig.~\ref{fig:mobility_finite_G_all_gate_gv1} shows the mobility and quantum mobility as a function of the electron density.
By comparing Figs.~\ref{fig:mobility_finite_G_all_gate} and \ref{fig:mobility_finite_G_all_gate_gv1}, it is evident that the disorder parameters for $g=4$ are generally higher than those for $g=2$, due to the fact that screening increases monotonically as $g$ increases.
Table~\ref{table:nc_air_gv1} shows the corresponding results of MIT critical densities.
We see that $n_c$ and $n_{cq}$ are in general smaller than the $g=2$ results shown in Table~\ref{table:nc_air}, because the values of best-fit mobility and quantum mobility are higher.
\begin{table}[b]
\caption{Critical density $n_c$ and $n_{cq}$ in units of $10^{11}$ cm$^{-2}$. The results are obtained from $k_F l = 1$ and $k_F l_q = 1$ using the same impurity parameters as in Fig.~\ref{fig:mobility_finite_G_all_gate_gv1} calculated in the finite potential well approximation with gate screening and local field correction. RI (BI) indicates the dominant scattering mechanism at low densities. 5 nm (7 nm) means the calculation for a 5 nm (7 nm) QW.}
\begin{ruledtabular}
\begin{tabular}{c c c c c} 
 & RI (5 nm) & RI (7 nm) & BI (5 nm) & BI (7 nm)\\
$n_c$ (g=2) & 0.44 & 0.41 & 0.16 & 0.15  \\
$n_{cq}$ (g=2) & 0.68 & 0.63 & 0.15 & 0.14  \\
\end{tabular}
\end{ruledtabular}
%\vspace{-0.2 in}
\label{table:nc_air_gv1}
\end{table}

\medskip
%\bibliographystyle{apsrev4-1}
%\bibliography{reference.bib}

\begin{thebibliography}{88}
\bibitem [{RI_()}]{RI_integral}%
  \BibitemOpen
  \href@noop {} {}\bibinfo {note} {These results are also true even in the weak screening limit $q_{TF}/2k_F \ll 1$ where the density is high $g/16 \pi d^2 \ll g^3/16 \pi a_B^2 \ll n_e$, as long as $q_{TF} d \gg 1$ is satisfied.}\BibitemShut {Stop}%
\bibitem [{IR_()}]{IR_integral}%
  \BibitemOpen
  \href@noop {} {}\bibinfo {note} {These results are also true even in the weak screening limit $q_{TF}/2k_F \ll 1$ where the density is high $g/4 \pi \Lambda^2 \ll g^3/16 \pi a_B^2 \ll n_e$, as long as $q_{TF} \Lambda \gg 1$ is satisfied.}\BibitemShut {Stop}%
\end{thebibliography}
%\medskip

\end{document}